\newif\ifhasappendix
\hasappendixfalse
\documentclass[sigconf,nonacm=true]{acmart}
\usepackage{tikz}
\usepackage{wrapfig}
\usepackage{multirow}
\usepackage{amsmath}
\usepackage{booktabs}
\usepackage{makecell} 
\usepackage{algorithm}
\usepackage{enumitem}
\usepackage[noend]{algorithmic}

\newcommand{\algorithmicoutput}{\textbf{output}}
\newcommand{\OUTPUT}{\item[\algorithmicoutput]}
\AtBeginDocument{%
  }

\copyrightyear{2025}
\acmYear{2025}
\setcopyright{cc}
\setcctype{by}
\acmConference[SC '25]{The International Conference for High Performance Computing, Networking, Storage and Analysis}{November 16--21, 2025}{St Louis, MO, USA}
\acmBooktitle{The International Conference for High Performance Computing, Networking, Storage and Analysis (SC '25), November 16--21, 2025, St Louis, MO, USA}
\acmDOI{10.1145/3712285.3759802}
\acmISBN{979-8-4007-1466-5/2025/11}

\begin{document}
\pagestyle{plain}

\title{BurstEngine: an Efficient Distributed Framework for Training Transformers on Extremely Long Sequences of over 1M Tokens}


\def\thu{Department of Computer Science and Technology, Tsinghua University, Beijing, China}
\def\bupt{Beijing University of Posts and Telecommunications, Beijing, China}
\def\ailab{Shanghai Artificial Intelligence Laboratory, Shanghai, China.}
\def\addr{\city{}\country{}}

\author{Ao Sun}
\authornote{indicates equal contribution.}
\email{maydomine@bupt.edu.cn}
\orcid{0009-0002-3631-1233}
\affiliation{%
  \institution{\bupt}
  \addr
}

\author{Weilin Zhao}
\authornotemark[1]
\email{zwl23@mails.tsinghua.edu.cn}
\affiliation{%
  \institution{\thu}
  \addr
}

\author{Xu Han}
\authornote{indicates corresponding authors.}
\email{han-xu@tsinghua.edu.cn}
\affiliation{%
  \institution{\thu}
  \addr
}

\author{Cheng Yang}
\authornotemark[2]
\email{yangcheng@bupt.edu.cn}
\affiliation{%
 \institution{\bupt}
 \addr
}

\author{Zhiyuan Liu}
\affiliation{
 \institution{\thu}
 \addr
}

\author{Chuan Shi}
\affiliation{%
  \institution{\bupt}
  \addr
}

\author{Mao\allowbreak song Sun}
\affiliation{%
  \institution{\thu}
  \addr
}

\renewcommand{\shortauthors}{Sun et al.}

\begin{abstract}

Existing methods for training LLMs on long-sequence data, such as Tensor Parallelism and Context Parallelism, exhibit low Model FLOPs Utilization as sequence lengths and number of GPUs increase, especially when sequence lengths exceed 1M tokens. To address these challenges, we propose BurstEngine, an efficient framework designed to train LLMs on long-sequence data. BurstEngine introduces BurstAttention, an optimized distributed attention with lower communication cost than RingAttention. BurstAttention leverages topology-aware ring communication to fully utilize network bandwidth and incorporates fine-grained communication-computation overlap. Furthermore, BurstEngine introduces sequence-level selective checkpointing and fuses the language modeling head with the loss function to reduce memory cost. Additionally, BurstEngine introduces workload balance optimization for various types of attention masking. By integrating these optimizations, BurstEngine achieves a $1.2\times$ speedup with much lower memory overhead than the state-of-the-art baselines when training LLMs on extremely long sequences of over 1M tokens. We have made our code publicly available on GitHub: \url{https://github.com/thunlp/BurstEngine}.

\end{abstract}





\maketitle

\section{Introduction}

Transformers~\cite{vaswani2017attention} have emerged as the dominant architecture for large language models (LLMs)~\cite{brown2020language,touvron2023llama2,dubey2024llama} and large multimodal models (LMMs)~\cite{dai2023instructblip,zhu2024minigpt,chen2024internvl}. 
However, training Transformer-based models on long sequences faces two challenges: the memory consumption associated with storing large intermediate states and the quadratic cost of the attention mechanism with respect to sequence lengths.
Considering that long-sequence training is crucial for LLMs and LMMs to generate extensive outputs such as documents, code, and videos, many efforts have been made to address the challenges associated with long-sequence training. 

\begin{figure}[t]
  \centering \includegraphics[width=0.9\linewidth]{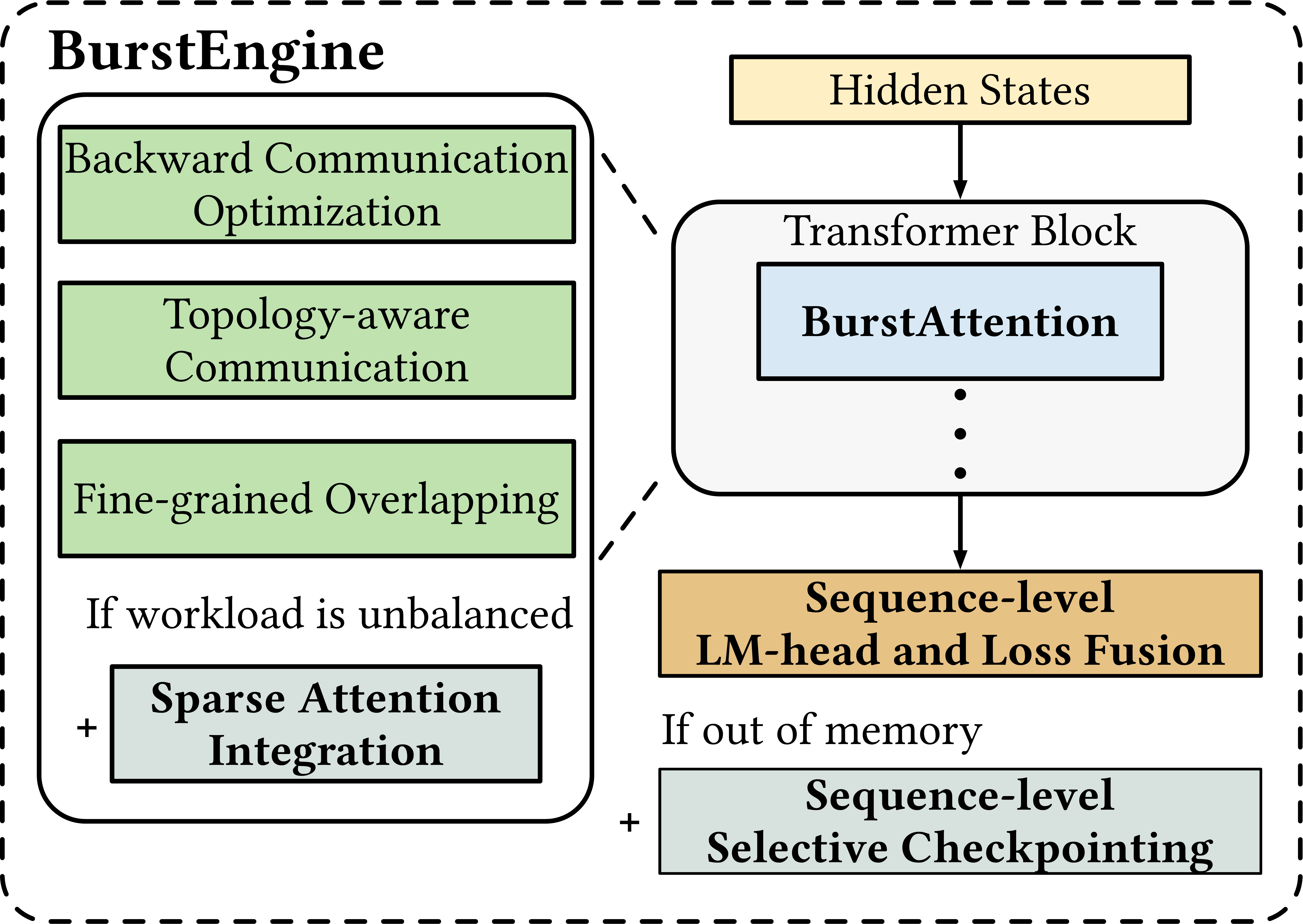}
  \caption{The overview of BurstEngine's main optimizations.} \label{fig:overview}
\end{figure}


These efforts are approached from two perspectives: one focuses on improving the efficiency of individual devices (e.g., single GPU) in processing long sequences, while the other leverages distributed systems (e.g., multiple GPUs) to handle long sequences. 
On the one hand, typical works such as FlashAttention~\cite{dao2022flashattention,dao2023flashattention} employ memory-friendly online softmax~\cite{milakov2018online}, enabling Transformer-based models to efficiently process sequences of 32K tokens using a single device. Other works, such as gradient checkpointing\cite{chen2016training, narayanan2021efficient}, employ recomputation to restore intermediate states rather than storing them, thus improving memory efficiency for long sequences. On the other hand, Context Parallelism (e.g., RingAttention)~\cite{gu2024loongtrain,liu2023ring}, Tensor Parallelism~\cite{narayanan2021efficient, korthikanti2023reducing}, and Head Parallelism (e.g., DeepSpeed-Ulysses)~\cite{jacobs2023deepspeed} are three typical approaches to leverage distributed clusters to train models on longer sequences. Recent LLMs and LMMs need to handle sequences beyond 128K tokens, making parallelism methods essential for training long-sequence Transformer-based models~\cite{abdin2024phi,dubey2024llama,yang2024qwen2,hui2024qwen2}. In this paper, we propose ``BurstEngine'', an efficient framework specifically designed to train Transformer-based LLMs and LMMs on extremely long sequences of over 1M tokens. As illustrated in Figure~\ref{fig:overview}, BurstEngine considers long-sequence optimizations from the perspective of the entire Transformer, especially handling the issues of attention for processing long sequences, including four parts: (1) \textbf{BurstAttention}, (2) \textbf{Sequence-level Selective Checkpointing}, (3) \textbf{Sequence-level Fusion of Language Modeling (LM) Head and Loss Function}, and (4) \textbf{Sparse Attention Integration}. 

\textbf{BurstAttention} is a highly efficient distributed attention implementation. As illustrated in Figure~\ref{fig:overview}, BurstAttention introduces three key optimizations: (1) Backward Communication Optimization, which reduces nearly 25\% of communication cost in the backward pass compared to existing efficient context parallelism by rewriting the backward pass of attention modules in a communication-efficient way. (2) Topology-aware ring communication, which splits communication into intra-node and inter-node communication, and thus fully takes advantage of the bandwidth of different networks to reduce communication cost. (3) Fine-grained communication-computation overlap, which designs a specialized double buffer to better overlap computation and communication. BurstAttention significantly enhances Transformers' efficiency for long sequences, but long sequence bring high memory consumption due to storing intermediate states. To address this challenge, we propose sequence-level selective checkpointing and sequence-level fusion of LM head and loss function.

\textbf{Sequence-level Selective Checkpointing} is a novel gradient recomputation scheme specialized for attention modules, which optimizes the trade-off between memory consumption and computational overhead at the sequence level. Unlike traditional approaches~\cite{chen2016training} that store or recompute entire sequences, it selectively checkpoints the former part of a sequence, storing the latter part and recomputing the former during backward passes. This approach can significantly reduce memory consumption while maintaining limited computational cost.

\textbf{Sequence-level Fusion of LM Head and Loss Function} can reduce the memory consumption of the Language Model (LM) head by fusing the LM head with the cross-entropy loss function to reduce the memory consumption of storing massive intermediate states. Furthermore, we fuse the forward and backward passes of the LM head and cross-entropy loss function to avoid recomputing related intermediate states.

\textbf{Sparse Attention Integration} enables BurstEngine to flexibly and efficiently incorporate BurstAttention with a variety of sparse patterns, including causal masking, sliding-window masking, dilated masking, and other block-wise sparse patterns.

We evaluate BurstEngine on a series of settings with up to 64$\times$ GPUs. The experimental results show that BurstEngine achieves a 1.2$\times$ speedup compared to the state-of-the-art method on extremely long sequences of over 1M tokens. Additionally, BurstEngine saves 26.4\% of memory compared to most memory-efficient baselines.

In summary, we make the following contributions: (1) We propose BurstAttention, a novel distributed attention with backward communication optimization, topology-aware ring communication pattern, and fine-grained overlapping. (2) We introduce a set of novel optimizations, including sequence-level selective checkpointing and sequence-level fusion of LM head and loss function, along with sparse attention integration. (3) We build BurstEngine, an implementation of the proposed framework with over 35K lines of Python, C++, and CUDA code, which achieves the state-of-the-art performance on training LLMs and LMMs on extremely long sequences of over 1M tokens.

\section{Preliminary}
\subsection{Preliminary of Transformers}
A Transformer consists of multiple blocks, each with an attention module and a feed-forward network (FFN), along with a language modeling (LM) head mapping the final block's output to vocabulary space. In this section, we introduce the attention module and the LM head in detail as they are key bottlenecks in long-context training.

\textit{\textbf{Transformer Attention Module.}} Given a sequence containing $N$ tokens as input, whose embeddings are denoted as $\mathbf{X} \in \mathbb{R}^{N \times d}$, the attention module can be formalized as
\begin{equation}
\begin{aligned}
\label{eq:attention}
\mathbf{Q} = \mathbf{X}\mathbf{W}_Q, \mathbf{K} = \mathbf{X}\mathbf{W}_K, \mathbf{V} = \mathbf{X}\mathbf{W}_V, &\\
\mathbf{S} = \frac{\mathbf{Q}\mathbf{K}^\top}{\sqrt{d}},
\mathbf{P} = \text{Softmax}(\mathbf{S}),
\mathbf{O} = \mathbf{P} \mathbf{V},
\mathbf{Y} = & \mathbf{O}\mathbf{W}_{\text{attn}},
\end{aligned}
\end{equation}
where $\mathbf{W}_Q, \mathbf{W}_K, \mathbf{W}_V \in \mathbb{R}^{d \times d}$ map $\mathbf{X}$ to $\mathbf{Q}, \mathbf{K}, \mathbf{V} \in \mathbb{R}^{N \times d}$, the query, key, and value embeddings.
$\mathbf{S} \in \mathbb{R}^{N \times N}$ and $\mathbf{P} \in \mathbb{R}^{N \times N}$ indicate the attention scores and the attention probabilities, respectively. 
$\mathbf{O} \in \mathbb{R}^{N \times d}$ is the sum of value embeddings weighted by query-key similarities, mapped to $\mathbf{Y} \in \mathbb{R}^{N \times d}$ via $\mathbf{W}_{\text{attn}} \in \mathbb{R}^{d \times d}$.
Most LLMs and LMMs use multi-head attention, where each head follows Eq.~(\ref{eq:attention}) and outputs are concatenated. For simplicity, we omit multi-head details as the number of heads does not affect our optimizations and conclusions, and denote the process as $\mathbf{Y} = \text{ATTN}(\mathbf{X})$. Since our efficiency methods apply broadly to various attention patterns, we refrain from detailed discussions on these patterns. 

\textit{\textbf{Transformer Block and LM head.}} A Transformer block $\mathbf{Y} = \text{Transformer}(\mathbf{X})$ is given as
\begin{equation}
\mathbf{H} = \text{ATTN}(\mathbf{X})+\mathbf{X}, \mathbf{Y} = \text{FFN}(\mathbf{H})+\mathbf{H}.
\end{equation}
After stacking $M$ Transformer blocks, we can build an LLM or LMM to generate tokens 
\begin{equation}
\begin{aligned}
\label{eq:lmhead}
\text{LM}(\text{Trans}&\text{former}_M(\cdots\text{Transformer}_1(\mathbf{X}))),
\end{aligned}
\end{equation}
where $\text{LM}(\mathbf{X})= \text{Softmax}(\mathbf{X}\mathbf{W}_{\text{vocab}}^\top)$ is the language modeling (LM) head that maps the outputs of a Transformer to the probability distribution, $\mathbf{W}_{\text{vocab}} \in \mathbb{R}^{v \times d}$ is the vocabulary embeddings, and $v$ is the size of the vocabulary set. For training LLMs and LMMs, the outputs of the LM head are used to calculate loss. For more details of Transformers, see \cite{vaswani2017attention} and surveys \cite{han2021pre,lin2022survey,han2022survey,khan2022transformers}.
\subsection{Challenges of Processing Long Sequences in Transformers}
\label{seq:issues}

As we mentioned before, the attention module and the LM head are key bottlenecks in training Transformers on long sequences.
In this section, we explain why and how they become bottlenecks in training Transformers on long sequences.

\setlength{\columnsep}{10pt}
\setlength{\intextsep}{10pt}
\begin{wrapfigure}[21]{c}[0pt]{0.2\textwidth}
    \centering
    \includegraphics[width=0.2\textwidth]{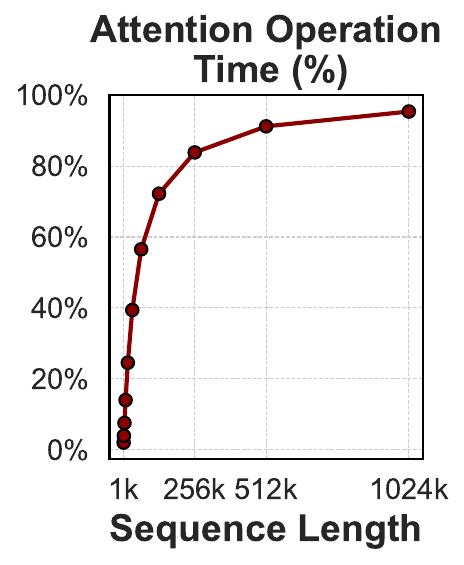}
  \vspace{-1em}
    \caption{The proportion of time spent by attention modules during the end-to-end training process of a Transformer-based model (7B parameters).}
    \label{fig:attn_scale}
  \vspace{-1em}
\end{wrapfigure}

(1) \textbf{Challenges in Attention Module.} 
The attention module exhibits quadratic complexity with respect to sequence lengths, making handling long sequences challenging in practice. As shown in Figure~\ref{fig:attn_scale}, attention modules have become the main bottleneck in training Transformer-based models on long sequences of over 128K tokens. Note that recent LLMs and LMMs are required to handle sequences of over 1M tokens. In this case, we have to use distributed clusters to make Transformers efficiently process long sequences.

(2) \textbf{Challenges of Storing Intermediate States.} Training Transformer models requires computing loss functions through forward passes and obtaining parameter gradients through backward passes. As sequence length increases, memory required to store intermediate states during the forward pass prior to performing the backward pass becomes a significant bottleneck. To address this, efficient memory optimization solutions are essential, especially for those GPUs with limited storage capacity but sufficient computing power.

(3) \textbf{Challenges in Language Modeling Head.} When computing the LM head, $\text{LM}(\mathbf{X})= \text{Softmax}(\mathbf{X}\mathbf{W}_{\text{vocab}}^\top)$, large memory is required to store the states $\mathbf{X}\mathbf{W}_{\text{vocab}}^\top \in \mathbb{R}^{N \times v}$ before the Softmax function. This memory consumption continues to grow as the sequence length increases, eventually reaching a point where it exceeds the capacity of a single GPU.

(4) \textbf{Challenges of Integrating Sparse Attention.} In LLMs and LMMs, attention modules are often coupled with sparse masking to control which keys and values participate in the attention mechanism. In other words, the probabilities in Eq.~(\ref{eq:attention}) are obtained by $\mathbf{P} = \text{Softmax}(\mathbf{M} \odot \mathbf{S})$, where $\mathbf{M} \in \mathbb{R}^{N\times N}$ is the masking matrix. For example, in LLMs, each token is required to only attend to the tokens preceding it in the sequence, which necessitates the use of a triangular masking matrix. Additionally, sparse attention mechanisms~\cite{winata2020lightweight,wang2020linformer} are often employed to speed up long-sequence processing. These sparse mechanisms rely on a sparse masking matrix during the attention process to reduce computational complexity. Obviously, the introduction of complex masking leads to an imbalanced workload, presenting challenges for using distributed clusters to process long sequences.

In subsequent sections, we will focus on introducing \textbf{how to address these challenges on distributed clusters and efficiently train Transformer-based models on long sequences}.

\begin{figure}[t]
\centering
\includegraphics[width=0.64\linewidth]{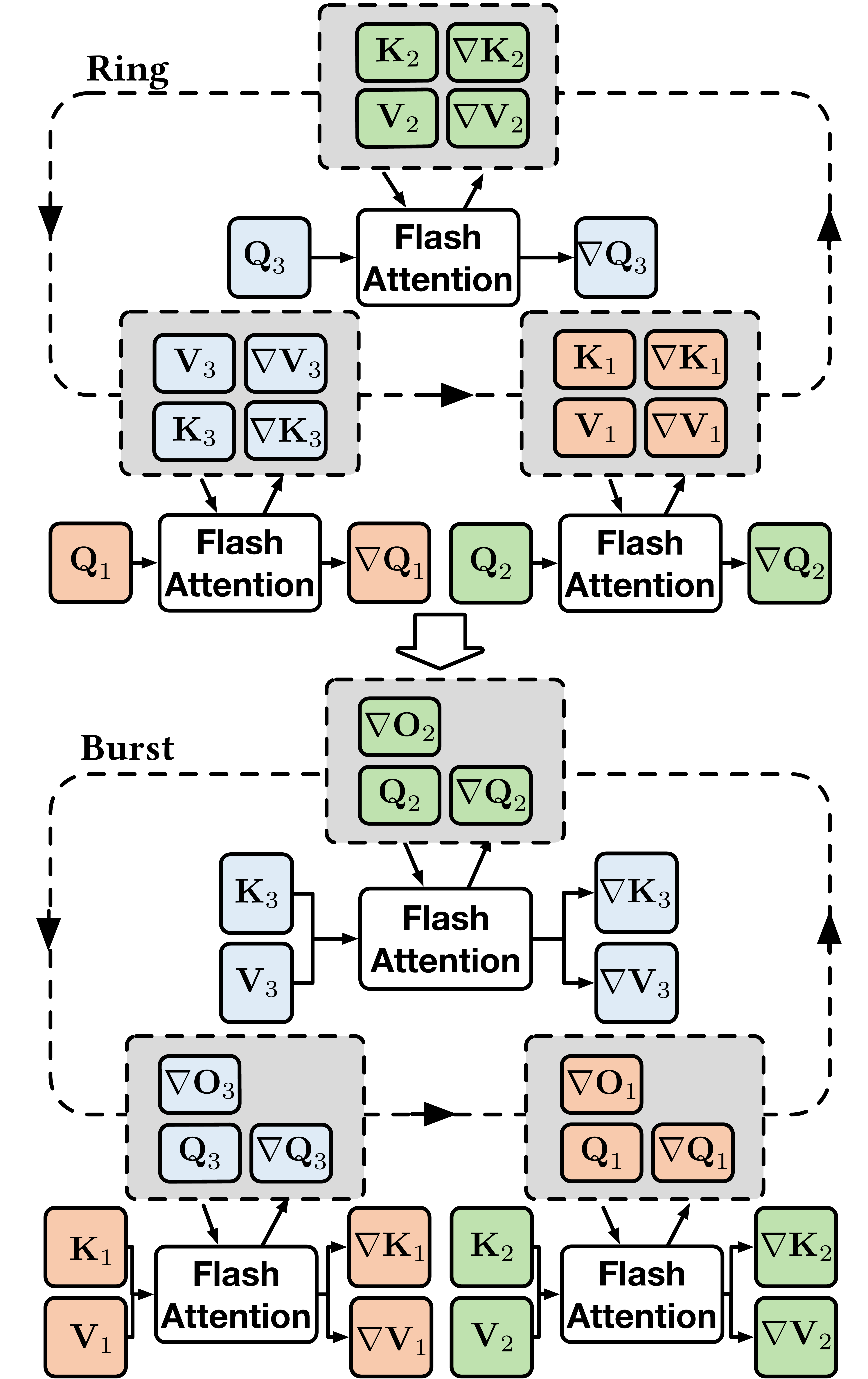}
\caption{The backward pass process of RingAttention and BurstAttention using $3$ GPUs.}\label{fig:backward_comm}
\end{figure}

\section{Overall Framework of BurstEngine}
\label{method}


\subsection{BurstAttention}
BurstAttention addresses the efficiency challenges encountered by the attention module during the training of Transformers on long-sequence data by leveraging a distributed cluster.
To facilitate introducing the details of BurstAttention, here we define a distributed cluster as a cluster built by several nodes, and each node contains several GPUs. Similar to the recent \textbf{competitive Context Parallelism method RingAttention}, after obtaining $\mathbf{Q}, \mathbf{K}, \mathbf{V} \in \mathbb{R}^{N \times d}$ in Eq.~(\ref{eq:attention}), BurstAttention divides these sequence embeddings into multiple partitions along the sequence dimension according to the number of GPUs in the cluster. Each GPU is assigned a query partition, a key partition, and a value partition. Formally, given the GPU number $G$ of the cluster, the $i$-th GPU is assigned $\mathbf{Q}_i, \mathbf{K}_i, \mathbf{V}_i \in \mathbb{R}^{\frac{N}{G} \times d}$. Since the parallelism efficiency primarily depends on minimizing communication costs while maximizing the overlap between communication and computation, BurstAttention introduces three optimizations to reduce communication cost and achieve better scalability and efficiency: backward communication optimization, topology-aware ring communication, and fine-grained overlapping of communication and computation. Next, we will first introduce how BurstAttention completes basic forward passes, and then introduce three optimizations in detail.

\textit{\textbf{Forward Pass of BurstAttention.}} BurstAttention formalizes the forward pass of an attention module into a multi-step process. At each step, the $i$-th GPU receives  $\mathbf{K}_j$ and $\mathbf{V}_j$ from its neighbor and performs local attention operations, and these local operations can be formalized as
\begin{equation} 
\label{eq:local-attention} 
\mathbf{S}_{i,j} = \frac{\mathbf{Q}_i\mathbf{K}_j^\top}{\sqrt{d}},  \mathbf{P}_{i,j} = \text{Softmax}(\mathbf{S}_{i,j}), \mathbf{O}_{i,j} = \mathbf{P}_{i,j}\mathbf{V}_j, 
\end{equation}
where $\mathbf{O}_{i,j} \in \mathbb{R}^{\frac{N}{G} \times d}$ indicates the local hidden states. $\mathbf{S}_{i,j} \in \mathbb{R}^{\frac{N}{G} \times \frac{N}{G}}$ and $\mathbf{P}_{i,j} \in \mathbb{R}^{\frac{N}{G} \times \frac{N}{G}}$ indicate the local attention scores and the local attention probabilities, respectively. 
After local attention operations, the $i$-th GPU sends $\mathbf{K}_j$ and $\mathbf{V}_j$ to its next neighbor for the next step. Within each single GPU, we adopt FlashAttention to efficiently complete local attention operations. For more details about FlashAttention, please refer to its related papers~\cite{dao2022flashattention,dao2023flashattention}.

This multi-step process continues until all $\mathbf{K}$ and $\mathbf{V}$ partitions have gone around the ring and completed all local attention operations. Generally, after obtaining all local states $\{\mathbf{O}_{i,j}\}_{i=1,j=1}^{G,G}$, we have to introduce additional communication and computation to aggregate these states into global states, as well as a lot of memory consumption to store these local states before aggregating them. To avoid incurring additional overhead, during the ring process, we introduce online softmax~\cite{milakov2018online} to progressively aggregate all local states $\{\mathbf{O}_{i,j}\}_{j=1}^{G}$ into the partitioned global states $\mathbf{O}_i\in\mathbb{R}^{\frac{N}{G}\times d}$, and finally map the concatenated partitioned global states $\{\mathbf{O}_i\}_{i=1}^{G}$ to the output embeddings $\mathbf{Y} \in \mathbb{R}^{N \times d}$.
Specifically, we reformulate the local operations in Eq.~(\ref{eq:local-attention}) as
\begin{equation}
  \begin{aligned}
  \label{eq:lseattn}
  \mathbf{S}_{i,j} = \frac{\mathbf{Q}_i\mathbf{K}_j^\top}{\sqrt{d}}, 
  \mathbf{Lse} = \text{LSE}(\mathbf{S}_{i,j}), \\
    \mathbf{P}_{i,j}=\exp\big(\mathbf{S}_{i,j} - \mathbf{Lse}\big), \mathbf{O}_{i,j} = \mathbf{P}_{i,j}\mathbf{V}_j. \\
    \end{aligned}
\end{equation}
where LogSumExp (LSE) function maps $\mathbb{R}^{\frac{N}{G} \times \frac{N}{G}}$ to $\mathbb{R}^{\frac{N}{G}}$ and given 
\begin{equation}
[\mathbf{Y}]_i = \log\sum_{j=1}^{\frac{N}{G}}\exp([\mathbf{X}]_{i:j}), 
\end{equation}
where $[\mathbf{Y}]_i$ is the $i$-th element of the vector $\mathbf{Y}$, $[\mathbf{X}]_{i:j}$ is the element located in the $i$-th row and the $j$-th column of the matrix $\mathbf{X}$. With Eq.~(\ref{eq:lseattn}), we can avoid storing $\{\mathbf{S}_{i,j}\}_{i=1,j=1}^{G,G}, \{\mathbf{P}_{i,j}\}_{i=1,j=1}^{G,G}, \{\mathbf{O}_{i,j}\}_{i=1,j=1}^{G,G}$ to obtain $\mathbf{O}_i$. Note that, the subtraction of the matrix $\mathbf{S}_{i,j}$ and the vector $\mathbf{Lse}$ requires broadcasting the elements of $\mathbf{Lse}$ along the last dimension of $\mathbf{S}_{i,j}$, and we will not repeatedly emphasize this.

\begin{algorithm}[t]
\footnotesize
\begin{algorithmic}[1]
\caption{The backward pass of RingAttention}
\label{alg:ring_backward}
\REQUIRE{Matrices $\mathbf{Q}_i,\mathbf{K}_i, \mathbf{V}_i, \mathbf{O}_i,\mathbf{\nabla O}_i \in \mathbb{R}^{{\frac{N}{G}}\times d}$, $\mathbf{Lse}_i\in\mathbb{R}^{\frac{N}{G}}$ on the $i$-th GPU
\STATE Initialize $\mathbf{\nabla Q}_i,\mathbf{\nabla K}_i,\mathbf{\nabla V}_i=(0)_{{\frac{N}{G}}\times d}$
\STATE Put $\mathbf{K}_i, \mathbf{V}_{i}, \mathbf{\nabla K}_i, \mathbf{\nabla V}_i $ into communication ring
}
\FOR{$j=1$ to $G$}
    \STATE Conduct one step of the ring communication;
    \STATE Get $\mathbf{K}_j, \mathbf{V}_j, \mathbf{\nabla K}_j,\mathbf{\nabla V}_j$ from communication ring;
    \STATE $\mathbf{S}_{i,j} = \mathbf{Q}_i\mathbf{K}_j^\top$;
    \STATE $\mathbf{P}_{i,j} = \text{exp}(\mathbf{S}_{i,j} - \mathbf{Lse}_i)$;
    \STATE $\mathbf{\nabla V}_j = \mathbf{\nabla V}_j + \mathbf{P}_{i,j}^\top\mathbf{\nabla O}_i$;
    \STATE $\mathbf{\nabla P}_{i,j} = \mathbf{\nabla O}_i~\mathbf{V}_j^\top$;
    \STATE $\mathbf{D}_i = \text{rowsum}(\mathbf{\nabla O}_i \circ \mathbf{O}_i)$  
    \STATE $\mathbf{\nabla S}_{i,j} = \mathbf{P}_{i,j}\circ(\mathbf{\nabla P}_{i,j}-\mathbf{D}_i)$;
    \STATE $\mathbf{\nabla K}_j = \mathbf{\nabla K}_j + \mathbf{\nabla S}_{i,j}^\top \mathbf{Q}_i$;
    \STATE $\mathbf{\nabla Q}_i = \mathbf{\nabla Q}_i + \mathbf{\nabla S}_{i,j}~\mathbf{K}_j$ ;
   \STATE Put $\mathbf{K}_j, \mathbf{V}_{j}, \mathbf{\nabla K}_j, \mathbf{\nabla V}_j$ into communication ring; \COMMENT{4Nd}
\ENDFOR
\OUTPUT $\mathbf{\nabla Q}_i, \mathbf{\nabla K}_i, \mathbf{\nabla V}_i$;
\end{algorithmic}
\end{algorithm}

\begin{algorithm}[h]
\footnotesize
\begin{algorithmic}[1]
\caption{The backward pass of BurstAttention}
\label{alg:burst_backward}
\REQUIRE{Matrices $\mathbf{Q}_i,\mathbf{K}_i, \mathbf{V}_i, \mathbf{O}_i,\mathbf{\nabla O}_i \in \mathbb{R}^{{\frac{N}{G}}\times d}$, $\mathbf{Lse}_i\in\mathbb{R}^{\frac{N}{G}}$ on the $i$-th GPU
\STATE Initialize $\mathbf{\nabla Q}_i,\mathbf{\nabla K}_i,\mathbf{\nabla V}_i=(0)_{{\frac{N}{G}}\times d}$\
\STATE $D_i = \text{rowsum}(\mathbf{\nabla O}_i \circ \mathbf{O}_i)$ 
\STATE Put $\mathbf{Q}_i, \mathbf{\nabla Q}_i, \mathbf{\nabla O}_i, \mathbf{D}_i, \mathbf{Lse}_i$ into communication ring
}
\FOR{$j=1$ to $G$}
    \STATE Conduct one step of the ring communication;
    \STATE Get $\mathbf{Q}_j, \mathbf{\nabla Q}_j, \mathbf{\nabla O}_j, \mathbf{D}_j, \mathbf{Lse}_j$ from communication ring;
    \STATE $\mathbf{S}_{j,i} = \mathbf{Q}_j\mathbf{K}_i^\top$;
    \STATE $\mathbf{P}_{j,i} = \text{exp}(\mathbf{S}_{j,i} - \mathbf{Lse}_j)$;
    \STATE $\mathbf{\nabla V}_i = \mathbf{\nabla V}_i + \mathbf{P}_{j,i}^\top\mathbf{\nabla O}_j$;
    \STATE $\mathbf{\nabla P}_{j,i} = \mathbf{\nabla O}_j~\mathbf{V}_i^\top$;
    \STATE $\mathbf{\nabla S}_{j,i} = \mathbf{P}_{j,i}\circ(\mathbf{\nabla P}_{j,i}-\mathbf{D}_j)$;
    \STATE $\mathbf{\nabla K}_i = \mathbf{\nabla K}_i + \mathbf{\nabla S}_{j,i}^\top \mathbf{Q}_j$;
    \STATE $\mathbf{\nabla Q}_j = \mathbf{\nabla Q}_j + \mathbf{\nabla S}_{j,i}~\mathbf{K}_i$ ;
  \STATE Put $\mathbf{Q}_j, \mathbf{\nabla Q}_j, \mathbf{\nabla O}_j, \mathbf{D}_j, \mathbf{Lse}_j$ into communication ring; \COMMENT{3Nd + 2N}
\ENDFOR
\OUTPUT $\mathbf{\nabla Q}_i, \mathbf{\nabla K}_i, \mathbf{\nabla V}_i$;
\end{algorithmic}
\end{algorithm}

\textit{\textbf{Communication Optimization of Backward Pass.}} Given a sequence consisting of $N$ tokens, whether using RingAttention or BurstAttention, the $i$-th GPU has
$\mathbf{Q}_i, \mathbf{K}_i, \mathbf{V}_i \in \mathbb{R}^{\frac{N}{G} \times d}$
and the partitioned global states $\mathbf{O}_i \in \mathbb{R}^{\frac{N}{G} \times d}$ after finishing forward pass. The communication cost of forward pass is $2Nd$, because each GPU receives and sends $\{\mathbf{K}_{j}\}_{j=1}^{G}$ and $\{\mathbf{V}_{j}\}_{j=1}^{G}$ once. 

As shown in Figure~\ref{fig:backward_comm} and Algorithm~\ref{alg:ring_backward}, during the backward pass of RingAttention, $\mathbf{K}_{j}, \mathbf{V}_{j}$ and their corresponding gradients $\nabla\mathbf{K}_{j}, \nabla\mathbf{V}_{j}$ are passed around the ring. At each step, the $i$-th GPU receives $\mathbf{K}_j, \mathbf{V}_j, \nabla\mathbf{K}_{j}, \nabla\mathbf{V}_{j}$ from its previous neighbor, and then use $\mathbf{K}_j, \mathbf{V}_j$ and the locally stored $\mathbf{Q}_i, \mathbf{O}_i, \nabla\mathbf{O}_i$ to update $\nabla\mathbf{Q}_i, \nabla\mathbf{K}_{j}, \nabla\mathbf{V}_{j}$.
After updating gradients, the $i$-th GPU sends the received partitions $\mathbf{K}_j, \mathbf{V}_j$ and the updated gradients $\nabla\mathbf{K}_{j}, \nabla\mathbf{V}_{j}$ to its next neighbor for the next step. It is evident that for RingAttention, the communication cost during the backward pass amounts to $4Nd$ per GPU, doubling the cost incurred during the forward pass.

To reduce the communication cost of the backward pass, BurstAttention adopts a different strategy from RingAttention. Specifically, since $\mathbf{P}_i = \text{Softmax}(\mathbf{S}_i)$ and $\nabla\mathbf{P}_i = \nabla\mathbf{O}_i \mathbf{V}^\top$, we can get
\begin{equation} 
\begin{aligned}
  \nabla\mathbf{S}_i &= \mathbf{P}_i \circ \nabla\mathbf{P}_i - \mathbf{P}_i \nabla\mathbf{P}_i^\top \mathbf{P}_i = \mathbf{P}_i \circ \nabla\mathbf{P}_i - \mathbf{P}_i (\nabla\mathbf{O}_i \mathbf{V}^\top)^\top \mathbf{P}_i \\
  &= \mathbf{P}_i \circ \nabla\mathbf{P}_i - (\mathbf{P}_i \mathbf{V}) \nabla\mathbf{O}_i^\top \mathbf{P}_i  = \mathbf{P}_i \circ \nabla\mathbf{P}_i - \mathbf{O}_i \nabla \mathbf{O}_i^\top \mathbf{P}_i,
\end{aligned}
\end{equation}
where $\circ$ denotes element-wise multiplication. Given $\mathbf{D}_i=\mathbf{O}_i \nabla \mathbf{O}_i^\top$,
\begin{equation}
\nabla\mathbf{S}_i = \mathbf{P}_i \circ \nabla\mathbf{P}_i - \mathbf{D}_i \mathbf{P}_i.
\end{equation}
Based on the above derivation, BurstAttention stores $\mathbf{K}_i, \mathbf{V}_i, \nabla\mathbf{K}_i, \nabla\mathbf{V}_i$ on each GPU, and passes $\mathbf{Q}_j, \nabla\mathbf{Q}_j, \nabla\mathbf{O}_j, \mathbf{Lse}_{j}, \mathbf{D}_{j}$ around the ring. Through formula substitution, we can pass $\mathbf{D}_{j}$ instead of $\mathbf{O}_{j}$ around the ring, and the whole backward pass of BurstAttention is shown in Algorithm~\ref{alg:burst_backward}. Since the total size of $\{\mathbf{D}_{j}\}_{j=1}^{G}$ and $\{\mathbf{Lse}_{j}\}_{j=1}^{G}$ is $N$, we can see that the communication cost of backward pass is $3Nd + 2N$ for BurstAttention, nearly $25\%$ less than RingAttention's $4Nd$. In addition, the computation overhead of BurstAttention is also lower than that of other RingAttention since we do not need to calculate $\text{rowsum}(\mathbf{\nabla O}_i \circ \mathbf{O}_i) $ in each round.

\textit{\textbf{Topology-aware Ring Communication and Fine-grained Communication-Computation Overlapping.}} Traditional Context Parallelism~\cite{li2021sequence,liu2023ring,gu2024loongtrain} experiences serious performance degradation in multi-node settings because of the limited communication bandwidth of inter-node networks. To address this, DoubleRing Attention~\cite{gu2024loongtrain} partitions global ring communication into intra-node and inter-node communication. This approach enhances the utilization of inter-node network interface controllers (NICs) and benefits from the overlapping of heterogeneous communication. While DoubleRing Attention offers notable advantages by overlapping inter-node communication, intra-node communication, and computation during forward passes, it fails to overlap gradient communication during backward passes. This oversight inevitably leads to performance degradation as the cost of gradient communication grows. To overcome these limitations, we introduce the topology-aware ring communication with fine-grained overlapping.

\begin{figure}[t!]
  \centering
  \includegraphics[width=0.72\linewidth]{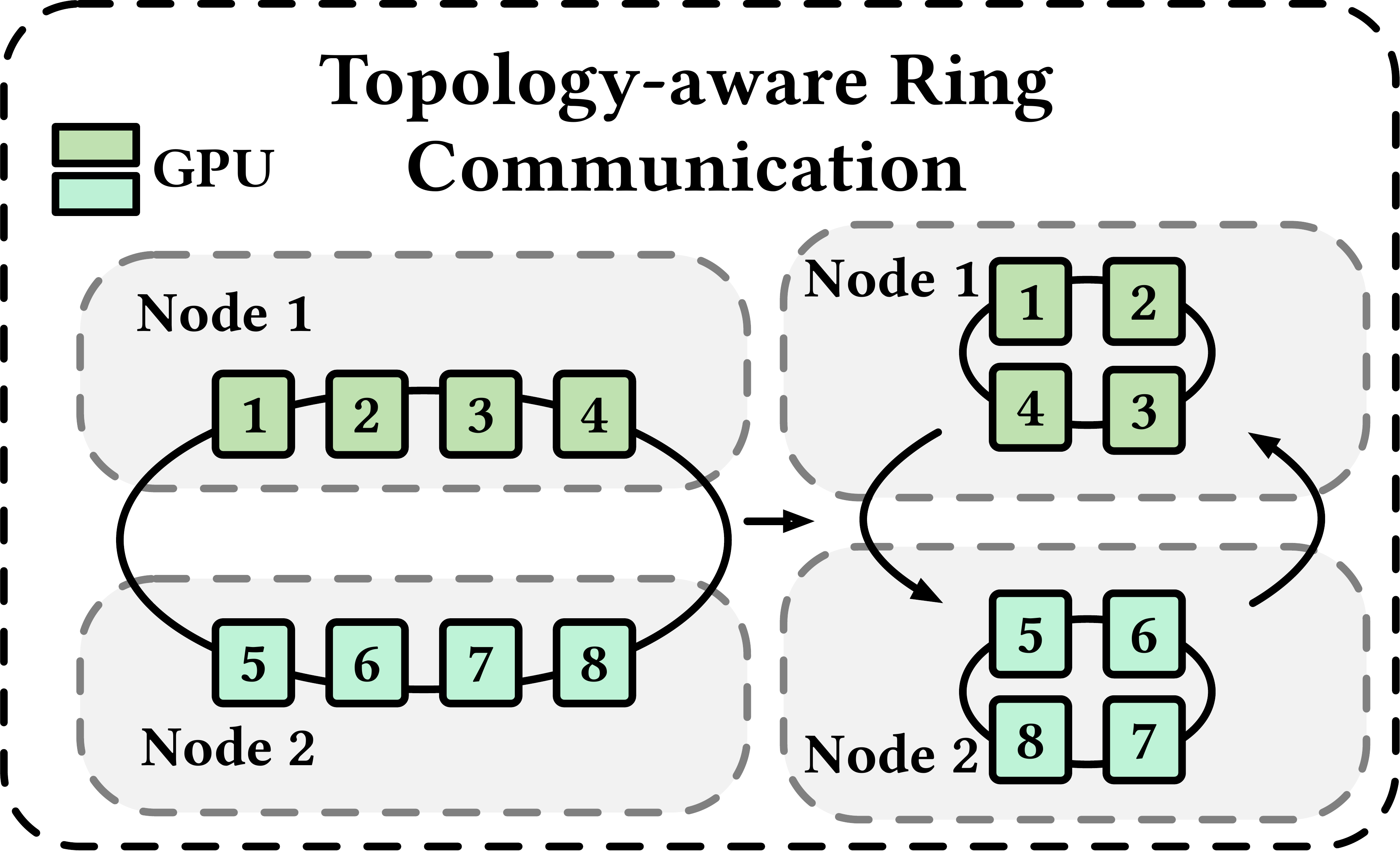}
  \caption{An example of topology-aware ring communication using $2\times4$ GPUs.}
  \label{fig:topo_ring}
\end{figure}

\begin{table}[H]
\caption{The communication time (forward and backward pass) of RingAttention, DoubleRingAttention, and BurstAttention. $\text{T}_{\text{intra}} = \text{Lat}_{\text{intra}} + \frac{P}{B_{\text{intra}}}$. $\text{T}_{\text{inter}} = \text{Lat}_{\text{inter}} + \frac{P}{B_{\text{inter}}}$. $\text{Lat}_{\text{intra}}$ and $\text{Lat}_{\text{inter}}$ are the latencies of intra-node communication and inter-node communication, respectively. $B_{\text{intra}}$ and $B_{\text{inter}}$ are the bandwidths of intra-node communication and inter-node communication, respectively. $P$ is the size of data to be communicated, $N_{\text{intra}}$ is the number of GPUs in the same node, $N_{\text{inter}}$ is the node number, and $G$ is total number of GPUs in all nodes.}
  \label{tab:communication_time}
\centering
\begin{tabular}{c|c}
\hline
\textbf{Method} & \textbf{Communication Time} \\ \hline
\multirow{1}{*}{RingAttention} & $6\max(N \cdot \text{T}_{\text{intra}},N \cdot \text{T}_{\text{inter}})$  \\ \hline
\multirow{2}{*}{DoubleRing} & $4 \max((N - N_{\text{inter}})\cdot \text{T}_{\text{intra}}, N_{\text{inter}}\cdot \text{T}_{\text{inter}})$ \\
                                          & $+ 2((N - N_{\text{inter}}) \cdot \text{T}_{\text{intra}} + N_{\text{inter}} \cdot \text{T}_{\text{inter}})$  \\ \hline
\multirow{1}{*}{BurstAttention} & $5 \max((N - N_{\text{inter}})\cdot \text{T}_{\text{intra}},$  $N_{\text{inter}} \cdot \text{T}_{\text{inter}})$  \\ \hline
\end{tabular}
\vspace{0.5em}
\end{table}

 As Figure~\ref{fig:topo_ring} shows, in a $2$-node cluster, where each node contains $4$ GPUs interconnected via NVLink, and the nodes are interconnected with the InfiniBand network, BurstAttention divides the global ring into two sub-rings, each responsible for communication among the GPUs within the same node. During each round of communication, GPUs within the same node perform a ring communication operation to exchange data. After all GPUs within the same node have completed their data exchanges ($4$ rounds intra-node communication in this case), GPUs in different nodes exchange data via the global ring. Considering there is commonly more than one NIC in distributed clusters for LLM training, the global ring communication can effectively utilize all available NICs' bandwidth and reduce inter-node communication cost significantly.

\begin{figure}[t!]
\centering
\includegraphics[width=\linewidth]{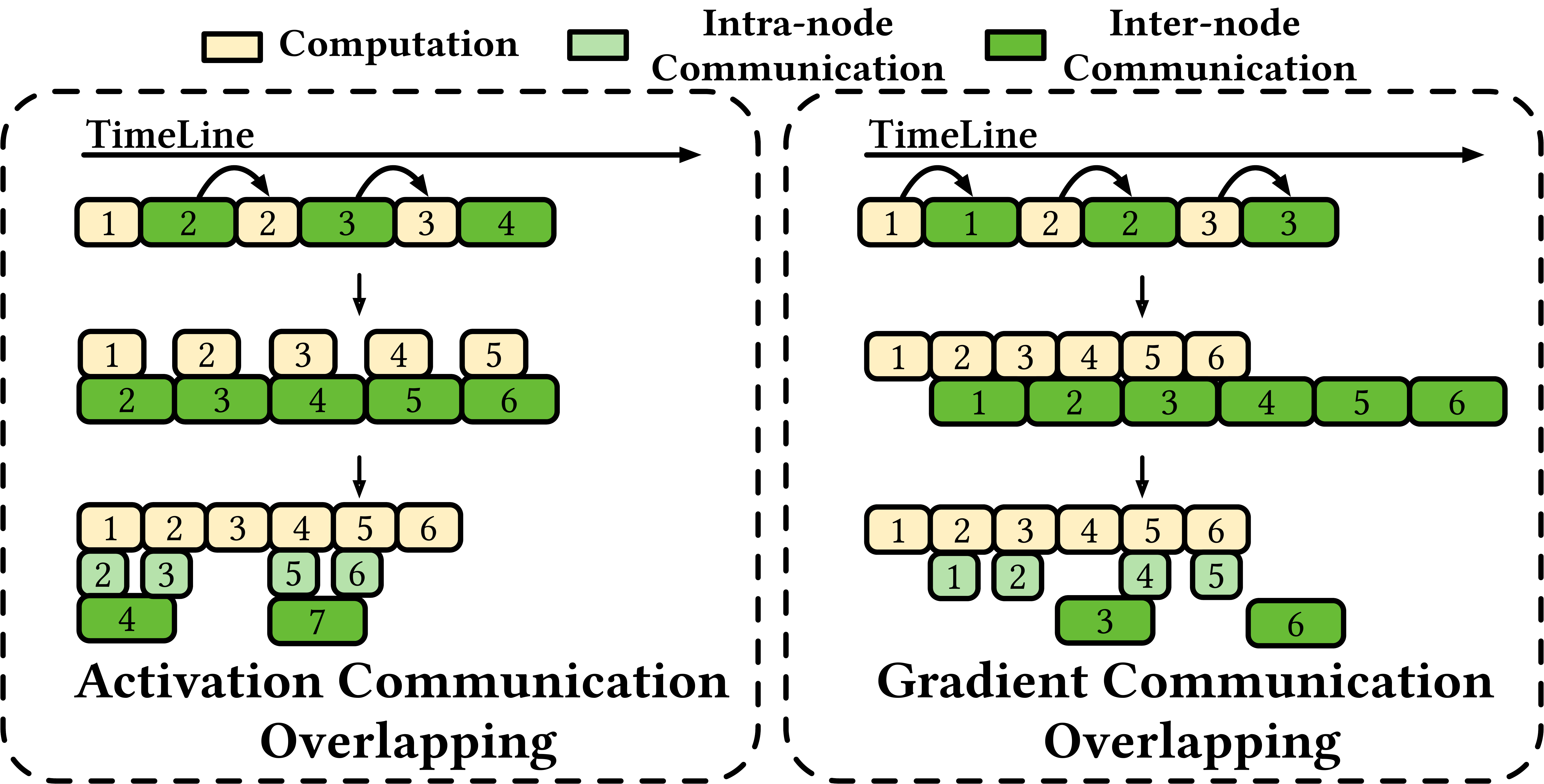}
\caption{The communication overlap for activations and gradients in BurstAttention.}\label{fig:overlap_timeline}
\end{figure}

Since intra-node and inter-node communication in Figure~\ref{fig:topo_ring} have separate railways in NVLink and InfiniBand networks, respectively, we can further overlap inter-node communication and intra-node communication. To achieve this, BurstAttention uses three dedicated buffers based on the topology: one for intra-node communication, one for inter-node communication, and one for computation. In each communication round, BurstAttention swaps the intra-node communication buffer with the computation buffer, allowing GPUs to obtain the data transmitted in the previous round for attention computation while continuing data exchange in this round. After all intra-node communication is complete, the inter-node communication buffer is swapped with the computation buffer, enabling GPUs to access data from remote nodes while initiating the next round of inter-node communication.

To achieve fine-grained overlapping between computation, intra-node communication, and inter-node communication, it is essential to properly manage the dependencies between communication and computation. Specifically, as illustrated in Figure~\ref{fig:overlap_timeline}, we categorize overlapping of BurstAttention into two types: 1) Activation (e.g., $\mathbf{K}, \mathbf{V}$) overlapping, where the first round of computation can be scheduled to execute concurrently with communication, as each round of communication is independent of the computation in the corresponding round; 2) Gradient (e.g., $\mathbf{\nabla Q}$) overlapping, where the first round of communication has a dependency for the computation of the corresponding round, thus we have to first compute the gradients and then communicate these gradients.



As illustrated in Figure~\ref{fig:overlap_timeline}, for activation overlapping, BurstAttention overlaps computation and communication by launching inter-node and intra-node communication threads simultaneously with computation threads. After one round of computation, BurstAttention waits for intra-node communication and launches another computation thread using received activations. After each device fully exchanges activations with other devices within the same node, it waits until inter-node communication is finished and then launches computation thread using received activations. The overlapping method for activations is not applicable for gradients since the first round of communication has a dependency on the gradient computation.
To solve this issue, as shown in Figure~\ref{fig:overlap_timeline}, BurstAttention delays intra-node communication and inter-node communication. Specifically, BurstAttention first performs one round of computation for warm-up. Then it initiates the intra-node communication and another round of computation, which gets rid of the dependency between later gradient communication and gradient computation. After all GPUs have exchanged the gradients with each other, BurstAttention launches the inter-node communication to exchange the gradients between different nodes. In this way, BurstAttention can almost fully overlap the computation, intra-node communication, and inter-node communication of both forward and backward passes. 

As shown in Table~\ref{tab:communication_time},  
BurstAttention has less communication time than RingAttention when $B_{\text{intra}}$ is larger than $B_{\text{inter}}$. By comparing the communication time of DoubleRingAttention and BurstAttention, we can find that BurstAttention not only has less communication cost due to backward communication optimization but also can save more communication time by overlapping inter-node communication and intra-node communication for gradients.

\begin{figure}[t]
  \begin{center}
    \includegraphics[width=1.0\linewidth]{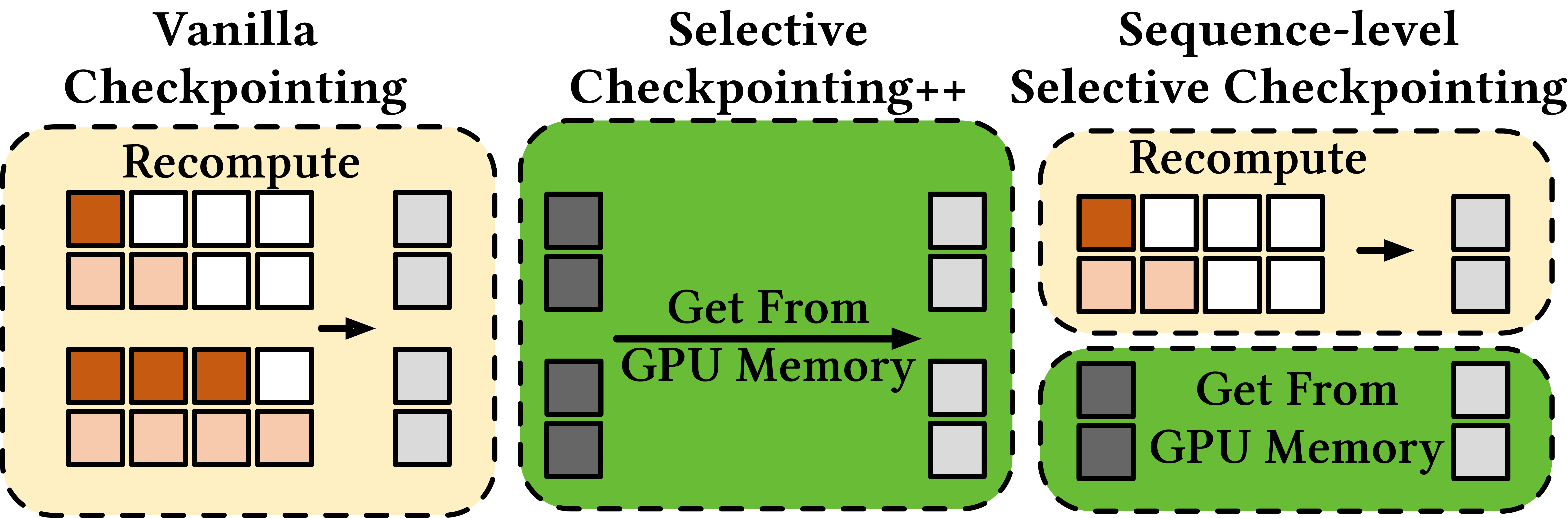}
  \end{center}
  \caption{The illustration of sequence-level selective checkpointing in BurstEngine.}\label{fig:recompute}
\end{figure}

\subsection{Sequence-level Selective Checkpointing}

\begin{figure}[t]
  \begin{center}
  \begin{minipage}{0.25\textwidth}
    \centering
    \includegraphics[width=\textwidth]{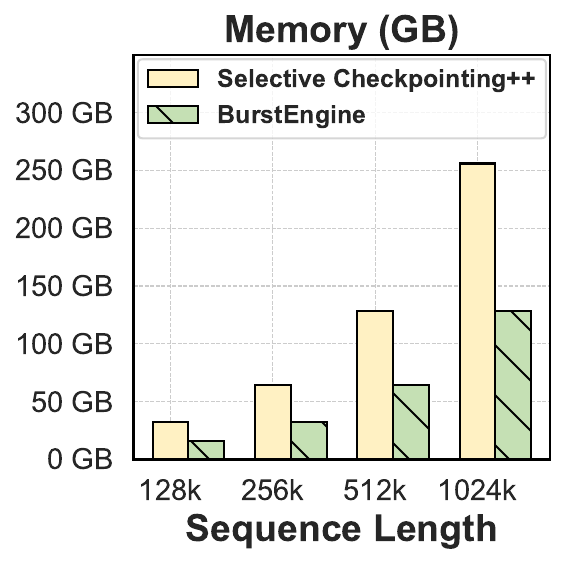}
    \caption{The total memory consumption of different gradient checkpointing strategies.}\label{fig:recompute_mem}
  \end{minipage}
  \hfill
  \begin{minipage}{0.21\textwidth}
    \centering
    \includegraphics[width=\textwidth]{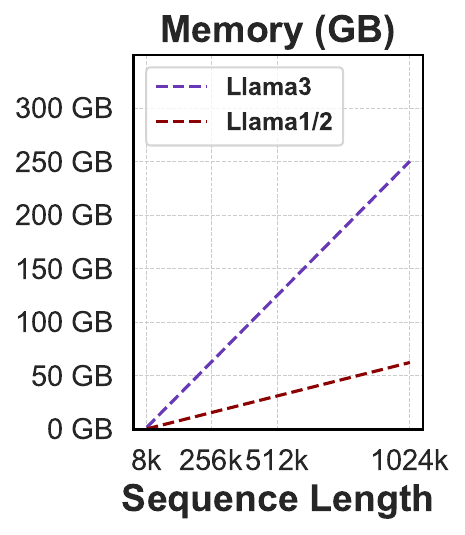}
    \caption{The total memory cost of LM head's logits for LLaMA-1/2 and LLaMA-3.}\label{fig:lm_head_mem}
  \end{minipage}
  \end{center}
\end{figure}

As the sequence length increases, the memory required to store intermediate states becomes a significant bottleneck. To address this, gradient checkpointing~\cite{chen2016training} is introduced to mitigate the high memory demands associated with storing these intermediate states. Generally, gradient checkpointing methods only store the input of each Transformer layer and get all intermediate states of each Transformer layer by using the input to do recomputation.


While gradient checkpointing methods reduce memory consumption, they increase computation overhead when combined with FlashAttention~\cite{dao2023flashattention}. This is mainly because FlashAttention~\cite{dao2023flashattention} fuses the Softmax process and matrix multiplications of the attention module, which means all intermediate states in attention would not be saved in forward passes and need to be recomputed in backward passes. To mitigate this issue, selective checkpointing++~\cite{li2024distflashattn, gu2024loongtrain} stores FlashAttention's output in memory and includes the outputs in a whitelist, avoiding recomputation and directly using the outputs stored in GPU memory. However, this method would introduce substantial memory consumption since the outputs of FlashAttention are large as the sequence length increases and the number of layers increases, as shown in Figure~\ref{fig:recompute_mem}.

To achieve a balance between memory consumption and computation overhead, we propose sequence-level selective checkpointing. As illustrated in Figure~\ref{fig:recompute}, the cost of recomputation for different sequence segments in the attention module is typically uneven, whereas the cost of storing activations remains the same in LLMs. 
Based on this observation, sequence-level selective checkpointing employs gradient checkpointing by dividing the sequence into two segments, storing the second segment with higher recomputation overhead, and only recomputing the first segment, as illustrated in Figure~\ref{fig:recompute}. In this manner, our method exhibits lower memory consumption with minimal additional computation overhead than other methods. As shown in Figure~\ref{fig:recompute_mem}, sequence-level selective checkpointing can reduce the memory consumption of gradient checkpointing by 50\% compared to selective gradient checkpointing++ while only introducing a slight decrease in end-to-end throughput.

\subsection{Sequence-level Fusion of Language Modeling Head and Loss Function}

In this section, we explain why the LM head becomes a key bottleneck in Transformers' long sequence training and introduce the details of the sequence-level fusion of the LM head and loss function.
First, we dive into the details of the LM head. Here, we refine Eq.~(\ref{eq:lmhead}) into
\begin{equation}
  \begin{aligned}
    \mathbf{Logits} &= \mathbf{H}_{\text{last}} \mathbf{W}_{\text{head}}^\top, \mathbf{P} = \text{Softmax}(\mathbf{Logits}), \\
    \mathcal{L} &= \text{Cross-Entropy}(\mathbf{P}, \mathbf{Y}),
  \label{eq:lm_head}
\end{aligned}
\end{equation}

where $\mathbf{H}_{\text{last}} \in \mathbb{R}^{N \times d}$ is the output embeddings of the last Transformer layer, $\mathbf{W}_{\text{head}} \in \mathbb{R}^{v \times d}$ is the vocabulary weight of the LM head, $\mathbf{P}$ is the probability distribution over the token vocabulary, and $\mathbf{Y}$ is the ground truth of the input.

To better support multiple languages and tasks, several LLMs have expanded their vocabulary sizes~\cite{dubey2024llama, hui2024qwen2}. Taking LLaMA~\cite{dubey2024llama,touvron2023llama,touvron2023llama2} as an example, the vocabulary size of the LLaMA series has grown from 32K in LLaMA-1 and
LLaMA-2~\cite{touvron2023llama, touvron2023llama2} to 128K in LLaMA-3~\cite{dubey2024llama}.
As the sequence length increases, the memory consumption of storing the outputs of the
LM head becomes a significant bottleneck, especially with a large vocabulary size. As shown in Figure~\ref{fig:lm_head_mem}, the memory consumption increases linearly with the sequence length, and the memory consumption of the LM head of LLaMA-3 is astonishingly high when dealing with long sequences. To address this issue, works like ~\cite{luo2024minisequence, wijmans2024cut} are proposed to split the hidden states and weights of the LM head into tiles, then fuse the LM head and cross-entropy at the tile level. In this way, there is no need to store the whole $\mathbf{P}$ matrix. During the backward pass, the above methods recompute the $\mathbf{P}$ matrix in the same way as the forward pass. However, these methods still introduce unnecessary computation overhead since they need to recompute the $\mathbf{P}$ in backward passes and would suffer from low training throughput.

\begin{figure}[t]
  \begin{center}
    \includegraphics[width=0.5\textwidth]{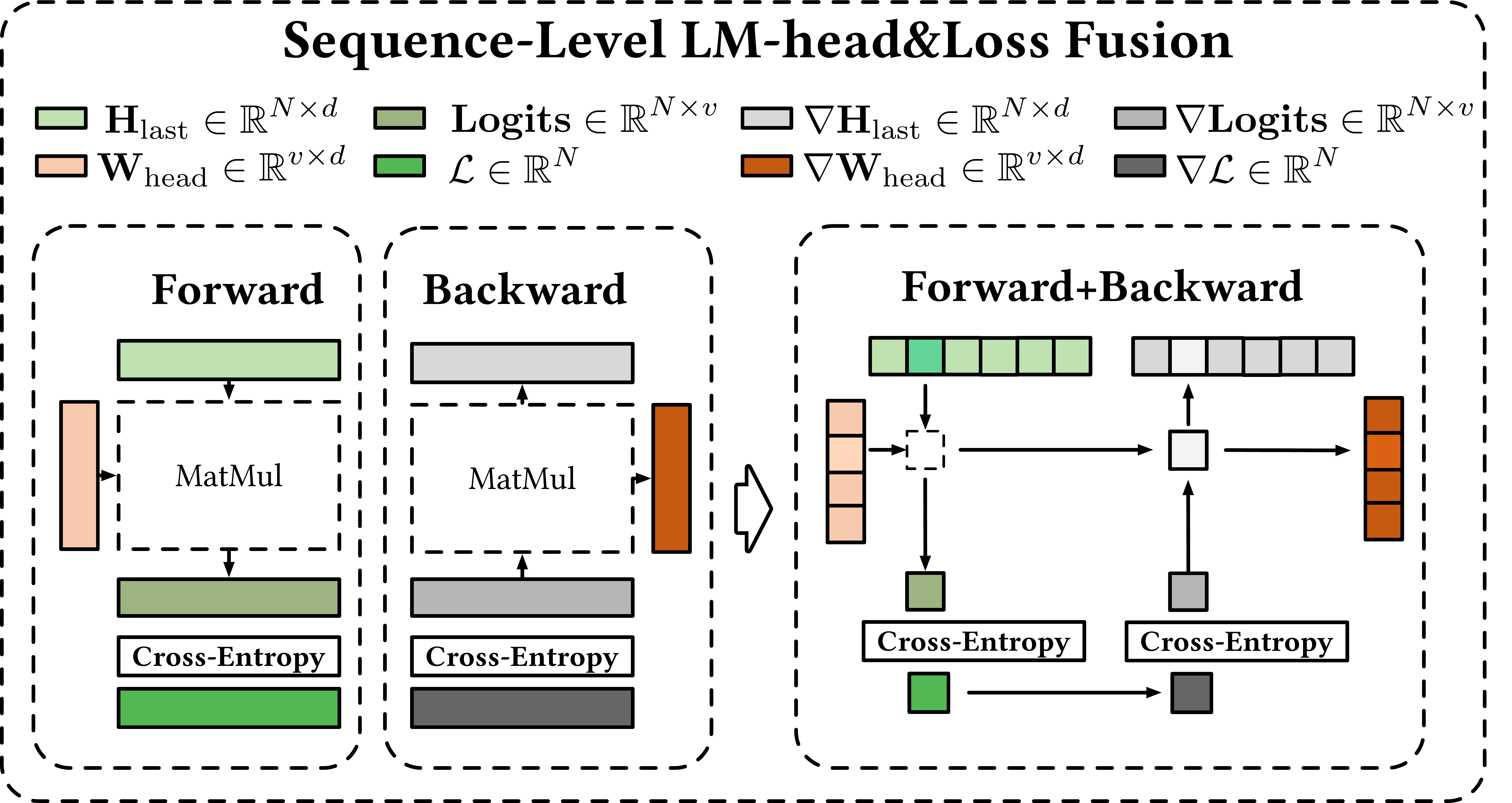}
  \end{center}
  \caption{The illustration of the sequence-level fusion of the LM head and loss function.}\label{fig:lm_head}
\end{figure}

To achieve better efficiency in training Transformers on long-sequence data, we introduce a sequence-level fusion of LM head and loss function without the need to recompute $\mathbf{P}$ and $\mathbf{Logits}$. As illustrated in Figure~\ref{fig:lm_head}, we propose to fuse the forward pass and backward pass in LM head and loss fusion. 
In detail, we divide $\mathbf{H}_{\text{last}}$ into tiles along the sequence dimension and $\mathbf{W}_{\text{head}}$ into tiles along the vocabulary dimension. During the forward pass, we loop over each tile of $\mathbf{H}_{\text{last}}$ and $\mathbf{W}_{\text{head}}$, and compute $\mathbf{Logits}$ and update $\mathbf{Lse}$ for each tile. After that, we compute the loss function for each tile of
$\mathbf{H}_{\text{last}}$ using $\mathbf{Lse}$ of $\mathbf{Logits}$. 
Then, we perform the backward pass immediately after the forward pass, without the need to
recompute $\mathbf{logits}$ and $\mathbf{P}$, which is the same idea as the online-softmax part of BurstAttention, as mentioned before. During the backward pass, we compute the $\nabla \mathbf{Logits}$, $\nabla \mathbf{H}_{\text{last}}$, $\nabla \mathbf{W}_{\text{head}}$ using tiles of $\mathbf{H}_{\text{last}}$ and $\mathbf{W}_{\text{head}}$. In this way, the sequence-level fusion of the LM head and loss function can reduce most memory consumption of storing the outputs of the LM head and avoid
unnecessary computation overhead in the backward pass, as depicted in Algorithm~\ref{alg:lmfuse}.

\begin{algorithm}[t]
\footnotesize
  \begin{algorithmic}[1]
    \caption{Sequence-level fusion of LM head and loss function}
    \label{alg:lmfuse}
  \REQUIRE{
    $\mathbf{H}_{\text{last}} \in \mathbb{R}^{N \times d}, \mathbf{W}_{\text{head}} \in \mathbb{R}^{v\times \text{d}}, \mathbf{Y} \in \mathbb{R} ^ {N}$ \\
    Block Size $B_s, B_v$ \\
    $\mathbf{W}_{\text{target}} = \mathbf{W}_{\text{head}}[\mathbf{Y}] $ \COMMENT{Index Operation, $\mathbf{W}_{\text{target}} \in \mathbb{R} ^ {N \times d}$}
  }

  \COMMENT{Divide $\mathbf{H}_{\text{last}}, \mathbf{W}_{\text{head}}$, $\mathbf{Y}$ into the tiles of the size $B_s \times d $, $B_v \times \text{d}$ and $B_s$} 
  \FOR{$ \text{blocks} \ \mathbf{H}_{i}\in\mathbb{R}^{B_s \times d}~\text{in}~\mathbf{H}_{\text{last}}$}
  \STATE $\mathbf{Lse}_{i} = -\infty$;
  \FOR{$ \text{blocks} \ \mathbf{W}_{j}\in\mathbb{R}^{\mathbf{B}_v\times d}~\text{in}~\textbf{W}_{\text{head}}$}
  \STATE $\mathbf{Logits}_{ij} = \mathbf{H}_{i} \mathbf{W}_{j}^\top$;
  \STATE $\mathbf{Lse}_{ij} = \log \sum \exp(\mathbf{Logits}_{ij})$;
  \STATE $\mathbf{Lse}_{i} = \log(\exp(\mathbf{Lse}_{i}) + \exp(\mathbf{Lse}_{ij}))$;
  \ENDFOR
  \STATE $\mathcal{L}_{i} = - \sum \mathbf{H}_{i} \cdot \mathbf{W}_{\text{target}} + \mathbf{Lse}_{i}$;

  \FOR {$\mathbf{blocks} \ \mathbf{W}_{j}\in\mathbb{R}^{B_v\times d}~\text{in}~\mathbf{W}_{\text{head}}$}
  \STATE $\nabla \mathbf{Logits}_{ij} = \exp(\mathbf{Logits}_{ij} - \mathbf{Lse}_{i})$;
  \STATE  $\mathbf{Index}_{j} = [j  B_v + 1, \cdots, (j+1) B_v]$;
  \STATE  $\mathbf{E} = 1$ if $\mathbf{Index}_{j} = \mathbf{Y}_{i}$, otherwise $0$; \COMMENT{$\mathbf{E} \ \in \mathbb{R}^{N}$}

  \STATE $\nabla \mathbf{H}_{i} \mathrel{{+}{=}} (\nabla \mathbf{Logits}_{ij} + \mathbf{E})\mathbf{W}_{j}^\top$.
  
  \STATE $\nabla \mathbf{W}_{j} \mathrel{{+}{=}} (\nabla \mathbf{Logits}_{ij} + \mathbf{E})^\top \mathbf{H}_{i}$.
  \ENDFOR

  \ENDFOR
  \end{algorithmic}
\end{algorithm}

\subsection{Sparse Attention Integration} 

In this section, we discuss how BurstAttention integrates with sparse
attention patterns, including common causal attention masking and other block-wise sparse attention masking. 

\textit{\textbf{Causal Attention.}} 
In the training process of most LLMs and LMMs, the goal is to predict the next token in the sequence.
In this task, the training objective is to maximize the log-likelihood of the next token given the previous tokens. To accomplish this, Transformer-based models commonly employ causal attention masking. For each token in the sequence, causal attention masking makes the token only attend to the key-value pairs of its preceding tokens. We can formulate this causal attention pattern as 
\begin{equation}
  \begin{aligned}
  \mathbf{O}_i = \text{Softmax}\left( \frac{\mathbf{Q}_i [\mathbf{K}_1,
  \mathbf{K}_2, \cdots, \mathbf{K}_{i}]^\top}{\sqrt{d}} \right)
  [\mathbf{V}_1, \mathbf{V}_2, \cdots, \mathbf{V}_{i}] 
  \label{eq:causal_attention} 
\end{aligned} 
\end{equation}
where $\mathbf{Q}_i$ is the query embedding of the $i$-th token, and $\mathbf{K}_j,
\mathbf{V}_j$ are the key and value embeddings of the $j$-th token. 

\begin{figure}[t]
  \begin{center}
    \includegraphics[width=0.38\textwidth]{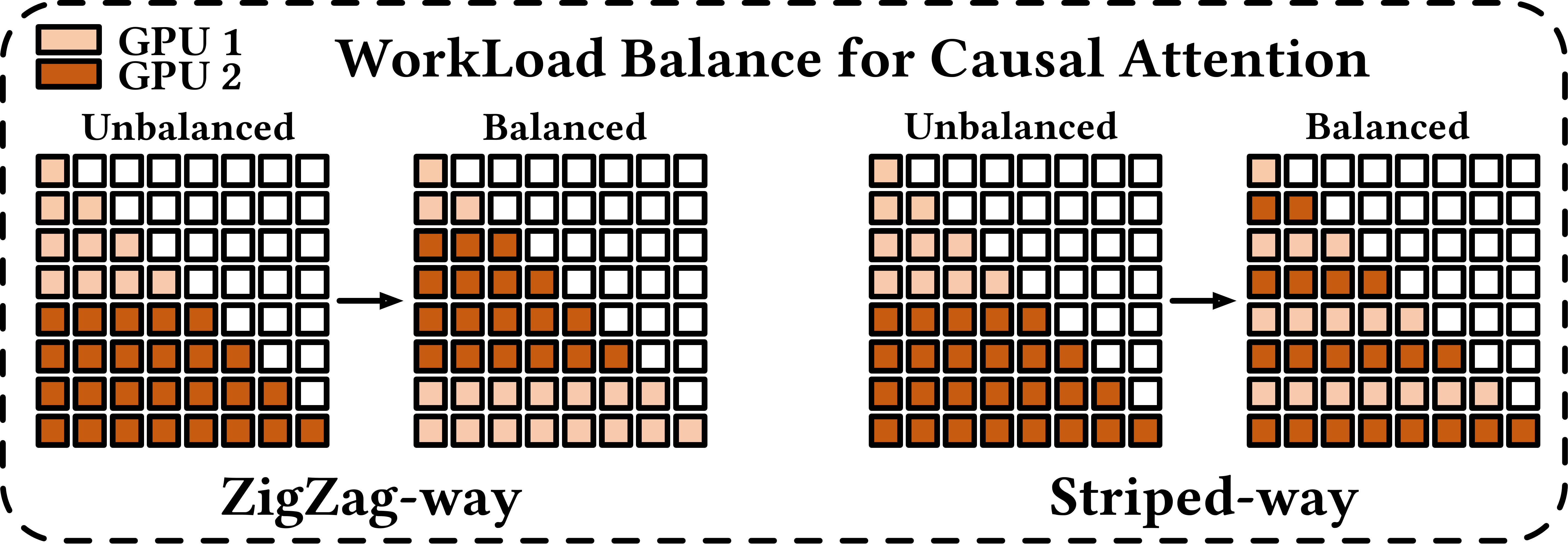}
  \end{center}
  \caption{Two types of workload balance methods for distributed causal attention.}\label{fig:causal_workload}
\end{figure}

To balance the workload of causal attention in Context Parallelism~\cite{gu2024loongtrain,liu2023ring}, two typical methods~\cite{brandon2023striped,nvidia2023cp,zhuzilin2023ring} have been proposed: striped-way workload balance and zigzag-way workload balance, as illustrated in Figure~\ref{fig:causal_workload}. Next, we will introduce how BurstAttention integrates these two kinds of methods to balance workloads.

For zigzag-way workload balance, the input sequence is initially partitioned along the sequence dimension across multiple devices. Given a sequence containing $N$ tokens $\{x_1, \ldots, x_n\}$ and $G$ devices, the sequence is divided into $2G$ subsequences, where the subsequence length is $P=\frac{N}{2G}$. 
For the $i$-th device ($1\leq i \leq  G$), it obtains two subsequences, one in the front and one in the back: $S^1_{i}$ and $S^2_{i}$, 
\begin{equation}
\begin{aligned}
  S^1_i &= \left \{ x_k \mid k \in [(i-1)\times P+1, i \times P]  \right\},  \\
  S^2_i &= \left \{ x_k \mid k \in [n-i\times P+1, n-(i-1)\times P]  \right\}.
\label{eq:zigzag_split}
\end{aligned}
\end{equation}
After getting subsequences, each device first performs causal attention on the
two subsequences and then communicates with other devices to obtain the key-value. 
Based on the partition, the $i$-th device has a front query and a back query, after getting the front key-value and back key-value from other devices. When the $i$-th device has the query embeddings $\{\mathbf{Q}^{1}_{i}, \mathbf{Q}^{2}_{i}\}$ of $S^{1}_{i}, S^{2}_{i}$ and receives $\{\mathbf{K}^{1}_{j}, \mathbf{K}^{2}_{j}\}, \{\mathbf{V}^{1}_{j}, \mathbf{V}^{2}_{j}\}$ of the $j$-th device's $S^{1}_{j}, S^{2}_{j}$, we get 
\begin{equation}
  \mathbf{O}_{ij} = 
  \begin{cases} 
    \text{CausalATTN}(\{\mathbf{Q}^{1}_{i}, \mathbf{Q}^{2}_{i}\}, \{\mathbf{K}^{1}_{j}, \mathbf{K}^{2}_{j}\}, \{\mathbf{V}^{1}_{j}, \mathbf{V}^{2}_{j}\}) & \text{if } i = j, \\
    \text{FullATTN}(\mathbf{Q}^2_{i}, \{\mathbf{K}^{1}_{j}, \mathbf{K}^{2}_{j}\}, \{\mathbf{V}^{1}_{j}, \mathbf{V}^{2}_{j}\}) & \text{if } i < j, \\
    \text{FullATTN}(\{\mathbf{Q}^{1}_{i}, \mathbf{Q}^{2}_{i}\}, \mathbf{K}^1_{j}, \mathbf{V}^1_{j}) & \text{if } i > j.
  \end{cases}
  \label{eq:zig-attn}
\end{equation}

For striped-way workload balance, the input sequence is initially partitioned along the sequence dimension across multiple devices into $G$ subsequences, where the subsequence length is $P=\frac{N}{G}$
For the $i$-th device ($1\leq i \leq  G$), it obtains one subsequence as
\begin{equation}
  S_i = \left \{ x_k \mid k \in \left\{i+G \times m \mid m \in [0, P-1] \right\}  \right\}.
  \label{eq:striped_split}
\end{equation}
The striped-way workload balance method ensures that each device can perform causal attention on the same number of tokens. When the $i$-th device has the query embeddings $\mathbf{Q}_{i}$ of $S_{i}$ and receives the key-value embeddings $\mathbf{K}_{j}, \mathbf{V}_{j}$ of the $j$-th device's $S_{j}$, we can get 
\begin{equation}
  \label{eq:strip-attn}
\begin{aligned}
  \mathbf{Q}'_{i} &= \{\mathbf{Q}_{ik} \mid k \in [2, P] \}, \\
  \mathbf{K}'_{i} &= \{\mathbf{K}_{ik} \mid k \in [1, P-1] \}, \\
  \mathbf{V}'_{i} &= \{\mathbf{V}_{ik} \mid k \in [1, P-1] \}, \\
  \mathbf{O}_{ij} &= 
  \begin{cases} 
    \text{CausalATTN}(\mathbf{Q}_{i}, \mathbf{K}_{j}, \mathbf{V}_{j}) & \text{if } i >= j, \\
    \text{CausalATTN}(\mathbf{Q}'_{i}, \mathbf{K}'_{j}, \mathbf{V}'_{j}) & \text{if } i < j.
  \end{cases}
\end{aligned}
\end{equation}

By adopting a zigzag-way workload balance or striped-way workload balance, each device has exactly the same compute workload, and FlashAttention's optimization of causal attention can be further leveraged to achieve workload balance at the streaming processor level. For BurstEngine, both zigzag-way workload balance and striped-way workload balance can be integrated. From our pilot experiments, integrating BurstEngine and striped-way workload balance achieves better performance.

\begin{figure}[t]
  \begin{center}
    \includegraphics[width=0.38\textwidth]{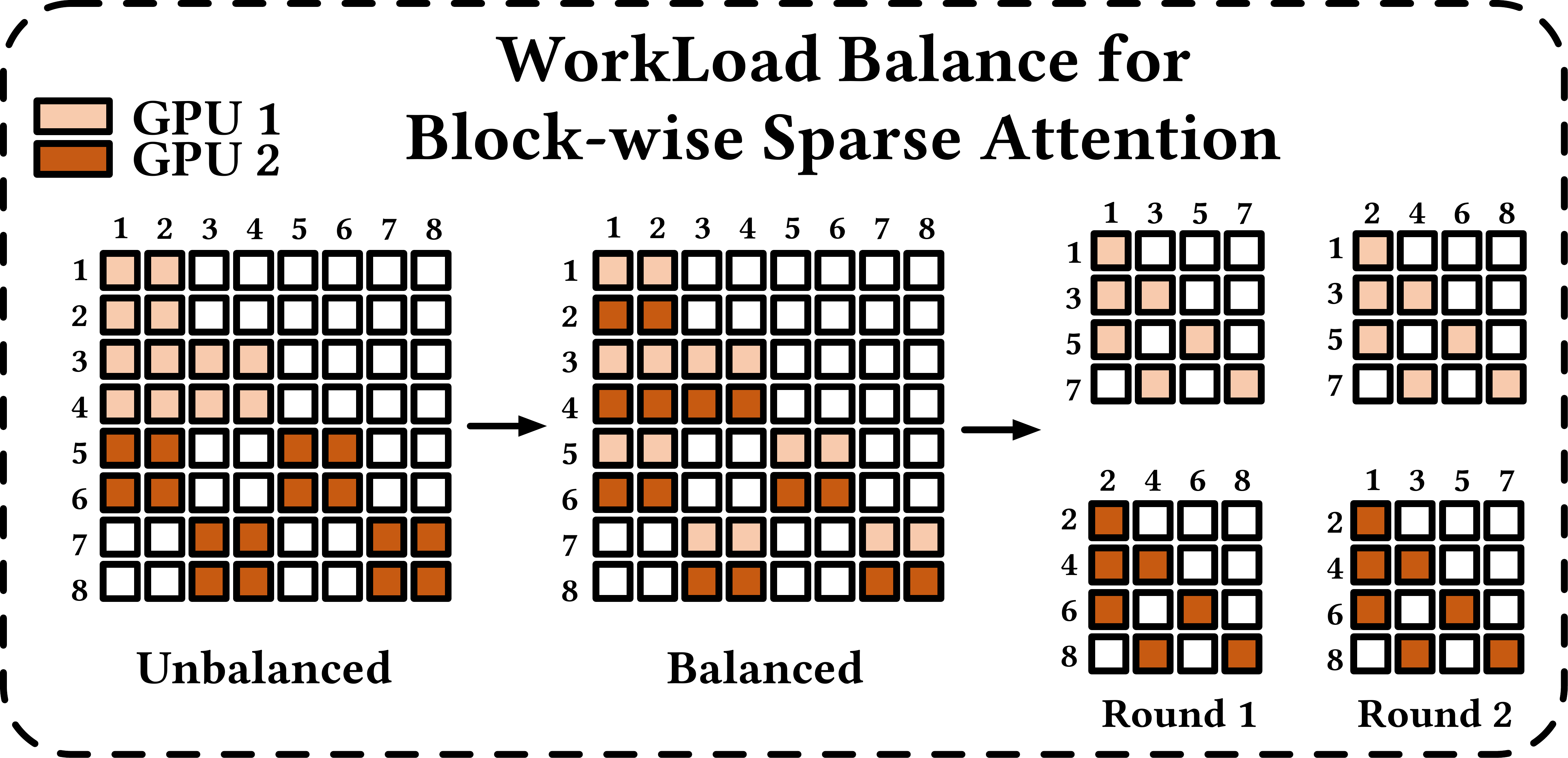}
  \end{center}
  \caption{Workload balance for block-wise sparse attention.}\label{fig:sparse_workload}
\end{figure}

\textit{\textbf{Block-wise Sparse Attention.}}
In addition to causal attention, block-wise sparse attention is another common sparse attention pattern. In block-wise sparse attention, the input sequence is divided into blocks, and each token only attends to tokens within the same block or some neighboring blocks. This pattern is widely used in reducing the computation and memory consumption of training Transformers~\cite{gray2017gpu,gale2023megablocks}. While sparse patterns are inherently difficult to integrate with conventional Context Parallelism methods due to extreme workload imbalance among workers, BurstEngine adopts a strategy similar to striped-way workload balance to balance the workload for block-wise sparse attention.  In block-wise sparse attention, we first divide the sequence into blocks along the sequence dimension, and the size of each block $\textbf{N}_{{\text{blk}}}$ needs to be a multiple of the number of devices $G$, which is a strict requirement for block-wise sparse attention workload balance. Then we define the block-masking matrix as $\textbf{M}_{\text{blk}} \in
\mathbb{R}^{\frac{\textbf{N}}{\textbf{N}_{\text{blk}}} \times
\frac{\textbf{N}}{\textbf{N}_\text{blk}}}$, where $\textbf{M}_{\text{blk}}[i,j]
= 1$ if all tokens in the $i$-th block can attend to the tokens in the $j$-th block.
Then, as shown in Figure~\ref{fig:sparse_workload}, we adopt a strategy similar to striped-way workload balance for block-wise sparse attention. With the strategy, each device can get exactly the same compute workload, and there is no unnecessary idle time for each device since the workload is balanced.


\section{Experiments}
\label{experiments}

\subsection{Experimental Settings}
\label{exp_setting}

\quad \textit{\textbf{Hardware Settings.}} We adopt two configurations: 32$\times$A800 and 64$\times$A800. For all configurations, each node is equipped with 8$\times$A800-SXM4-80GB GPUs linked via 400GB/s NVLink, 8 NVIDIA Mellanox HDR InfiniBand NICs(200Gb/s), and 128 CPU cores. 

\textit{\textbf{Model Sizes.}} In experiments, we mainly perform experiments on two settings of model sizes: LLaMA Transformer~\cite{touvron2023llama,touvron2023llama2,dubey2024llama} with 7 billion parameters (7B, 32 layers, 32 heads, 4096 dimensions, 32K vocabulary tokens) and with 14 billion parameters (14B, 40 layers, 40 heads, 5120 dimensions, 120K vocabulary tokens). 

\textit{\textbf{Training Settings.}} We adopt fully sharded data parallelism (FSDP)~\cite{zhao2023pytorch,rajbhandari2020zero} to partition model parameters across devices. All evaluations are conducted with gradient checkpointing~\cite{chen2016training} to achieve better training efficiency under limited memory. For 7B and 14B models, we respectively set the
sequence length to 2M and 1M on 32$\times$A800, and to 4M and 2M on 64$\times$A800.

\textit{\textbf{Implementation details}} For BurstEngine, we adopt the FSDP implementation from BMTrain~\cite{zeng2023bmtrain, wang2023h3t}, which achieves overlap of communication and computation at the Transformer-block level and supports optimizer offloading~\cite{ren2021zero}. Our Topology-aware ring communication is built on top of NCCL and uses multi-stream programming to overlap inter-node communication, intra-node
communication and computation. Additionally, we incorporate sequence-level fusion of LM head and loss fusion to reduce the memory consumption of storing the outputs of the LM head.

\textit{\textbf{Evaluation Metrics.}} For evaluation of training efficiency, we employ tokens per second per GPU (TGS) and model FLOPs utilization (MFU). TGS provides a direct measure of end-to-end training throughput, while MFU reflects the actual utilization of the GPU device. To evaluate memory consumption, we measure the peak memory allocated on each GPU for each method, since peak memory can directly evaluate the scalability of the method to larger models or longer sequences.

  \begin{figure}[t]
  \centering
      \includegraphics[width=0.43\textwidth]{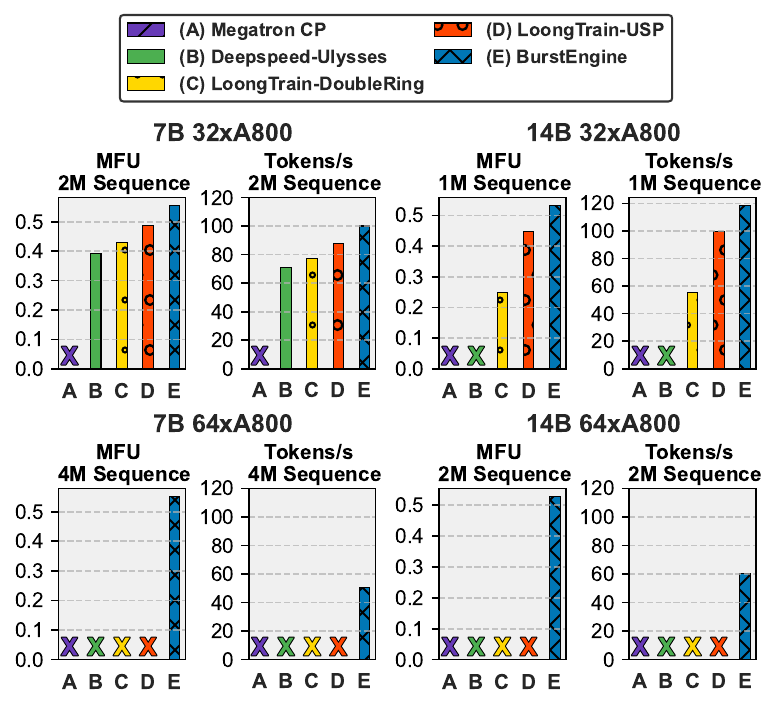}
      \caption{
        The end-to-end training throughput (TGS) of BurstEngine and other baselines.
      }
      \label{fig:main_exp}
  \end{figure}
  
  \begin{figure}[t]
  \centering
      \includegraphics[width=0.43\textwidth]{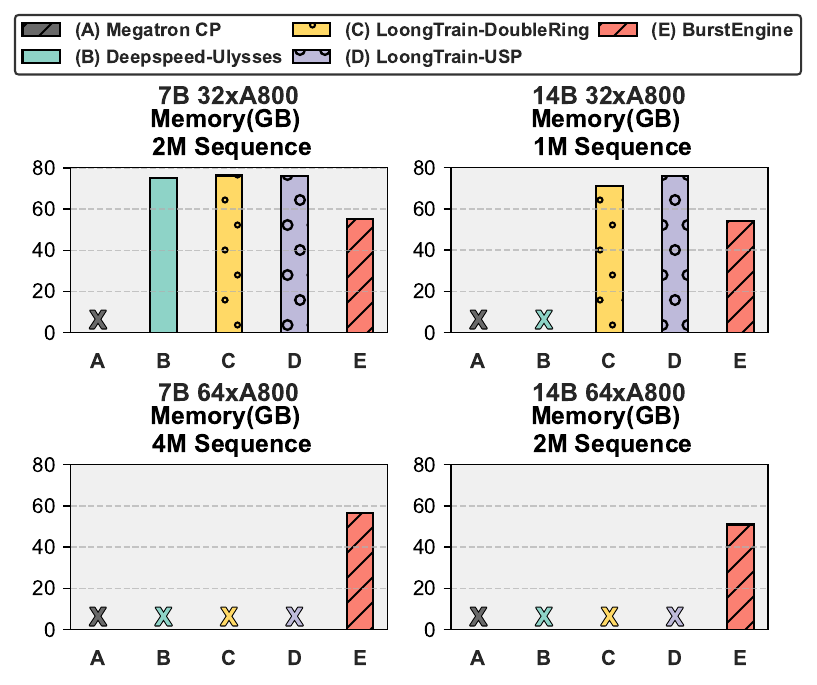}
      \caption{
       The peak memory usage (GB) per GPU in the end-to-end training of
     BurstEngine and other baselines. }
      \label{fig:main_exp_mem}
  \end{figure}

\textit{\textbf{Baselines.}}
We compare BurstEngine with the following baselines in our experiments:
(1) \textbf{Megatron Context Parallelism (CP)}: Megatron-LM's implementation~\cite{nvidia2023cp} for context parallelism, using RingAttention integrated with zigzag-way workload balance. (2) \textbf{DeepSpeed-Ulysses}: DeepSpeed's implementation~\cite{jacobs2023deepspeed} for head parallelism. (3) \textbf{LoongTrain-DoubleRing}: LoongTrain's implementation~\cite{gu2024loongtrain} for context parallelism, using DoubleRingAttention integrated with zigzag-way workload balance.
(4) \textbf{LoongTrain-USP}: LoongTrain's implementation for
Head-Context Hybrid Parallelism~\cite{gu2024loongtrain,fang2024unified}, achieving the state-of-the-art performance for long-sequence Transformer training, adopts head-first device placement and RingAttention with zigzag-way workload balance.

\subsection{Training Performance}

  \begin{figure}[t]
  \begin{center}
    \includegraphics[width=0.43\textwidth]{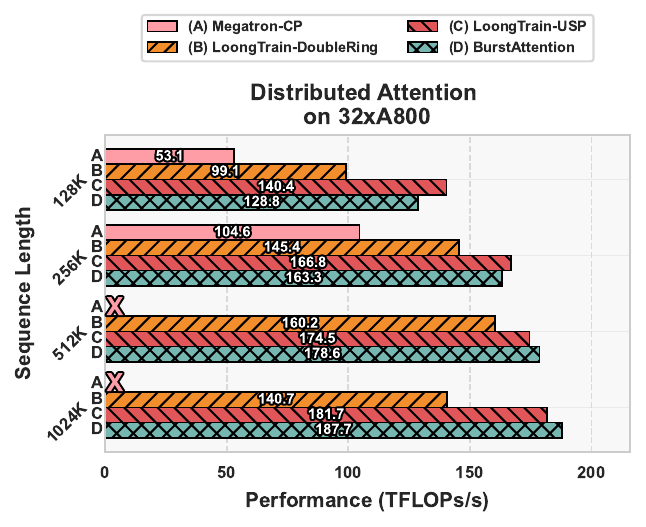}
  \end{center}
  \caption{Performance of different distributed attention implementations }\label{fig:attn_bench}
\end{figure}

  \quad\textit{\textbf{Throughput Performance.}} For comparisons of training efficiency, we evaluate the end-to-end training throughput of four baselines and BurstEngine on 7B and 14B models. As shown in Figure~\ref{fig:main_exp}, BurstEngine outperforms all baselines in terms of TGS and MFU on both 7B and 14B models.   BurstEngine achieves up to $1.19\times$/$1.15\times$ speedups on 7B/14B models, respectively, on 32$\times$A800 GPUs, compared to the state-of-the-art method LoongTrain-USP. In the 7B/14B model training under the 32$\times$A800 setting, Megatron-CP fails because of the out-of-memory issue, which is mainly due to the huge memory consumption of storing optimizer states and model weights since Megatron does not provide FSDP implementation and optimizer offloading. DeepSpeed-Ulysses has less memory consumption since it has FSDP and optimizer offloading, but still performs worse than LoongTrain-USP and BurstEngine because it can not overlap all-to-all communication with computation. LoongTrain-DoubleRing demonstrates superior throughput performance compared to DeepSpeed-Ulysses, leveraging its advanced capability to overlap communication with computation. However, it still underperforms compared to LoongTrain-USP, primarily because of the significant communication cost that remains unoptimized. 
  Though LoongTrain-USP has higher training throughput than other baselines, it still suffers from the communication cost of  Head Parallelism and RingAttention, respectively. BurstEngine achieves the best training efficiency (TGS and MFU), showing the efficiency of backward communication optimization and topology-aware ring communication with fine-grained overlap.
  
  \textit{\textbf{Memory Performance.}} In terms of memory performance, as illustrated in Figure~\ref{fig:main_exp_mem}, BurstEngine exhibits the lowest peak memory usage. The figure highlights that DeepSpeed-Ulysses, Loong-Train-DoubleRing, and LoongTrain-USP exhibit comparable memory consumption under the setting (7B model, 32$\times$A800), while BurstEngine saves $26.4\%$ memory compared to the best baseline. Under the setting (14B model, 32$\times$A800), DeepSpeed-Ulysses encounters an out-of-memory error due to its limitation on the number of model heads. LoongTrain-DoubleRing and LoongTrain-USP suffer from high memory consumption, which is mainly due to storing the outputs of the LM head. Under this setting, BurstEngine saves $24.2\%$ memory compared to the best baseline by adopting sequence-level fusion of LM head and loss function. Under the setting of 64$\times$A800, BurstEngine can support training a 7B model with a 4M
  sequence length and a 14B model with a 2M sequence length, which all baseline models fail to achieve. This is because BurstEngine exhibits nearly identical memory usage, indicating that BurstEngine achieves almost linear scaling with the device number along the sequence dimension.

\begin{table*}[t]
\centering
\caption{The ablation study of BurstEngine for using 32$\times$A800 to train a 14B model on the sequences of 1M tokens.}\label{tab:ablation}
\centering
\scalebox{0.92}{\begin{tabular}{cccccccc}
\toprule
\multicolumn{1}{c}{\makecell[c]{Backward\\Communication\\Optimization}} & 
\multicolumn{1}{c}{\makecell[c]{Topology-aware\\Ring Communication}} & 
\multicolumn{1}{c}{\makecell[c]{Sequence-level\\Fusion of \\LM head and Loss }} & 
\multicolumn{1}{c}{\makecell[c]{Sequence-level\\Selective\\Checkpointing}} & 
\multicolumn{1}{c}{\makecell[c]{Selective\\Checkpointing++}} & 
\multicolumn{1}{c}{\makecell[c]{MFU (\%)}} & 
\multicolumn{1}{c}{\makecell[c]{TGS}} & 
\multicolumn{1}{c}{\makecell[c]{Memory\\(GB)}} \\
\midrule
$\times$ & $\times$ & $\times$ & $\times$ & $\times$ & 36.75 & 83.79 & 48.47 \\
\checkmark & $\times$ & $\times$ & $\times$ & $\times$ & 38.37 & 87.48 & 49.31 \\
\checkmark & \checkmark & $\times$ & $\times$ & $\times$ & 41.69 & 95.06 & 48.97 \\
\checkmark & \checkmark & \checkmark & $\times$ & $\times$ & 41.58 & 94.81 &  \textbf{41.45}\\
\checkmark & \checkmark & \checkmark & \checkmark & $\times$ & 47.72 & 108.82 & 45.93 \\
\checkmark & \checkmark & \checkmark & $\times$ & $\checkmark$ & \textbf{51.68} & \textbf{117.83} & 53.91 \\
\bottomrule
\end{tabular}}
\end{table*}

\subsection{Attention Performance}
\label{attn_perf}
We evaluate the performance of BurstAttention using a 14B model's attention configuration on a setup of 32$\times$ A100 GPUs, comparing it with other RingAttention implementations such as Megatron's CP, LoongTrain's DoubleRingAttention, and USP. Given that the configuration involves 40 attention heads, DeepSpeed-Ulysses can not be applied in this scenario, as head parallelism is infeasible when the number of heads is not divisible by the number of GPUs. The experimental results, illustrated in Figure~\ref{fig:attn_bench}, highlight the efficiency of each method. From experimental results, Megatron's CP implementation encounters evaluation failures due to an out-of-memory issue when the sequence length exceeds 256k. Additionally, even in scenarios without memory issues, Megatron-CP performs poorly due to significant inter-node communication overhead. BurstAttention outperforms all other distributed attention implementations, achieving a $1.05\times$ speedup over LoongTrain's USP and a $1.33\times$ speedup over LoongTrain's DoubleRingAttention under 1M sequence setting. This superior performance can be attributed to BurstAttention's topology-aware ring communication strategy and its ability to finely overlap communication and computation.

At first glance, it might be a bit confusing why BurstAttention shows only marginal improvements compared to LoongTrain's USP, while BurstEngine achieves a $1.2\times$ speedup.
This is due to the difference between benchmarking attention alone and end-to-end training. When benchmarking attention alone, communication and computation can be overlapped perfectly. In end-to-end training, extra communication operations caused by FSDP result in huge communication costs and make perfectly overlapping impossible. In these cases, reducing communication cost leads to much bigger improvements in performance.

\subsection{Ablation study}

\quad \textit{\textbf{Main Optimization Strategies.}} We present an ablation study to assess the impact of individual optimization strategies in BurstEngine. The experiments are conducted using a $14$B model and sequences of $1$M tokens, evaluated on $32\times$A800. As shown in Table~\ref{tab:ablation}, we can observe how each optimization strategy contributes to the overall performance of BurstEngine.
Backward communication optimization brings approximately a $1.05\times$ speedup, while topology-aware ring communication and fine-grained communication-computation overlap contribute to an approximately $1.08\times$ speedup. Furthermore, sequence-level fusion of LM head and loss function can save $15.3\%$ memory compared to the baseline without introducing any performance degradation in training efficiency since there is no additional computation overhead. Sequence-level selective checkpointing can save another $14.8\%$ memory compared to the baseline and can achieve a $1.14\times$ speedup compared to the baseline with full checkpointing, positioning it as an optimal solution that balances memory and throughput. In the end, BurstEngine achieves a $1.4\times$ speedup and saves at least $15\%$ memory compared to the baseline without any optimization.

\textit{\textbf{Workload Balance for Sparse Attention.}} We present an ablation study to assess the impact of workload balance for sparse attention in BurstEngine. The experiments are conducted using a $14$B model and sequences of $1$M tokens, on $32\times$A800. We measure the training throughput of three sparse attention implementations:

\begin{table}[t]
    \caption{The throughput of integrating BurstEngine with different sparse attention masking strategies.}\label{tab:sparse_workload}
    \begin{center}
      \scalebox{0.95}{\begin{tabular}[c]{c|cc}
        \hline
        \multicolumn{1}{c|}{\textbf{Implementation}} & 
        \multicolumn{1}{c}{\textbf{TGS}} &
        \multicolumn{1}{c}{\textbf{Speedup}} \\
        \hline
        Attention Masking &  227.58 & $1.00 \times $ \\
        Causal Attention & 393.44 & $1.72\times$ \\
        SWA & 837.79 &  $3.68\times$ \\
        \hline
      \end{tabular}}
    \end{center}
  \end{table}
  
\begin{table}[t]
    \caption{The performance of BurstEngine across different nodes, with each node having 8$\times$A800.}\label{tab:inter_exp}
    \begin{center}
      \scalebox{0.85}{\begin{tabular}[c]{cc|ccc}
        \hline
        \multicolumn{1}{c}{\textbf{Nodes}} & 
        \multicolumn{1}{c|}{\textbf{Sequence }} & 
        \multicolumn{1}{c}{\textbf{MFU (\%)}} & 
        \multicolumn{1}{c}{\textbf{TGS}} & 
        \multicolumn{1}{c}{\textbf{Memory (GB)}} \\
        \hline
        2 & 0.5M &53.1 & 223.25 & 63.13 \\
        4 & 1M &53.2 & 118.36 & 53.96 \\
        8 & 2M &52.7 & 60.49 & 50.96 \\
        \hline
      \end{tabular}}
    \end{center}
  \end{table}
\begin{table}[t]
    \caption{The performance of BurstEngine across different context parallel size settings on 8$\times$A800.}\label{tab:intra_exp}
    \begin{center}
      \scalebox{0.85}{\begin{tabular}[c]{cc|ccc}
        \hline
        \multicolumn{1}{c}{\textbf{CP}} & 
        \multicolumn{1}{c|}{\textbf{Sequence}} & 
        \multicolumn{1}{c}{\textbf{MFU (\%)}} & 
        \multicolumn{1}{c}{\textbf{TGS}} & 
        \multicolumn{1}{c}{\textbf{Memory (GB)}} \\
        \hline
        1 & 32K & 47.34 & 1201.14 & 57.71 \\
        2 & 64K & 48.85 & 928.24 & 55.18 \\
        4 & 128K & 50.55 & 639.43 & 55.58 \\
        8 & 256K & 51.90 & 393.44 & 53.56 \\
        \hline
      \end{tabular}}
    \end{center}
  \end{table}

\textbf{BurstEngine w. Attention Masking} is the implementation of sparse attention without any workload optimization, which simply applies attention masking to restrict the attention range of each token and has the same computation overhead as full attention.

\textbf{BurstEngine w. Causal Attention} is the implementation of integrating BurstEngine with zigzag-way workload balance for causal attention, which can balance the workload of each device and avoid unnecessary communication and computation overhead.

\textbf{BurstEngine w. SWA} is the implementation of integrating BurstEngine with Sliding Window Attention (SWA). In this implementation, we adopt the block-wise sparse method, as illustrated in Figure~\ref{fig:sparse_workload}, to balance the workload of each device.

As shown in Table~\ref{tab:sparse_workload}, we can see that BurstEngine w. Causal Attention achieves a $1.72\times$ speedup compared to BurstEngine w. Attention Masking. As for the SWA pattern with a 32K sliding window size, BurstEngine achieves a $3.68\times$ speedup compared to the baseline without any workload balance strategy. While there still remains a gap in training efficiency compared to the theoretical optimization effect, BurstAttention has significantly reduced the idle occupancy caused by an imbalanced workload across devices. However, there remains potential for further optimization in communication patterns for sparse attention and single-device kernel optimization, which will be the focus of our future work.

  \subsection{Scalability Analysis}

  The scalability experiments are divided into two parts: 
  
  \textbf{Intra-node scalability}: In this experiment, we assess the training efficiency and memory consumption of BurstEngine using $8\times$A800 GPUs within a single node, evaluating its behavior under different context parallel sizes from $1$ to $8$. Besides, we enable optimizer offloading~\cite{ren2021zero} since the memory consumption of optimizer states is quite high as the world size is small.
  
  \textbf{Inter-node scalability}: We evaluate BurstEngine's training efficiency and memory consumption across different numbers of nodes from $2$ to $8$. Here, we disable optimizer offloading since optimizer states can be stored in each worker now. For all these settings, we set the sequence length per GPU to $32$K, and the total
  training sequence length is $\text{n}*32\text{k}$, in which $\text{n}$ is the number of GPUs. As shown in Table~\ref{tab:intra_exp}, BurstEngine's MFU exceeds $50\%$ in
  the single-node scenario with the context parallel size $\geq 4$  and the sequence length
  $\geq 128\text{K}$. In the $8\times$A800 setting, BurstEngine achieves a TGS of $393.44$
  tokens/s per GPU with a context parallel size of $8$ and a sequence length of
  $256$K. Meanwhile, BurstEngine's memory consumption remains stable as the context parallel size and the sequence length increase proportionally.

  Table~\ref{tab:inter_exp} shows BurstEngine can achieve an MFU of $52.7\%$ with $8$ nodes when processing sequences of $2$M tokens. The MFU remains stable and memory consumption as the node number and the sequence length increase, showing BurstEngine can scale when extending to multiple node settings. 

\section{Related Work}

To improve the efficiency of Transformers in processing longer sequences, several optimization techniques have been proposed.

Korthikanti et al.~\cite{korthikanti2023reducing} introduce selective activation recomputation, which avoids storing attention Softmax logits during the forward pass and then recomputes these logits during the backward pass. Rabe et al.~\cite{rabe2021self} formalize attention modules at the block level, and assign each thread block on the device to handle the attention computation of a subsequence, further reducing temporary memory usage and achieving logarithmic memory complexity with respect to sequence length. Dao et al.~\cite{dao2022flashattention} develop FlashAttention, a CUDA-based implementation that utilizes the high-speed I/O capabilities of SRAM to offer even greater performance improvements. While these works significantly reduce the memory consumption of attention modules for processing long sequences, they still face limitations due to the performance constraint of individual devices and bring huge computational costs as the sequence length grows.

To overcome this, some efforts have been made to utilize distributed clusters. Adopting general parallelism strategies~\cite{valiant1990bridging,huang2019gpipe,rajbhandari2020zero,ren2021zero} is the most straightforward, especially using Tensor Parallelism~\cite{narayanan2021efficient,korthikanti2023reducing}. Besides, Context Parallelism methods like RingAttention~\cite{li2021sequence,liu2023ring,gu2024loongtrain, fang2024unified} and Head Parallelism~\cite{jacobs2023deepspeed} like DeepSpeed-Ulysses have also been proposed, which distribute the attention computation across multiple devices along the sequence dimension or head dimension. Meanwhile, to better balance the recomputation cost of FlashAttention, Li et al.~\cite{li2024distflashattn} propose the selective checkpointing++, which stores FlashAttention's output and avoids recomputation in the backward pass. Other work, such as ~\cite{luo2024minisequence, wijmans2024cut} focus on optimizing the memory consumption of the LM head and propose adopting ideas similar to FlashAttention for the LM head and cross-entropy. 

While these works can support efficient training when sequence length grows to 128k, 256k, or even 512k, they still suffer from high memory consumption and performance degradation when sequence length extends to 1M or even longer. Moreover, existing methods poorly integrate with sparse attention, especially as sequences grow extremely long. Thereby, we propose BurstEngine, a system that builds upon these optimizations and further reduces the communication, memory, and computation overhead through our specific optimizations.

\section{Conclusion}

BurstEngine presents a novel and efficient framework for training transformer-based models on long-sequence data. By introducing BurstAttention, sequence-level selective checkpointing, sequence-level fusion of language modeling head and loss function, and sparse attention integration, BurstEngine addresses the scalability and efficiency challenges posed by long-sequence training. These optimizations achieve a $1.2\times$ speedup over the state-of-the-art baseline while reducing memory consumption by $26.4\%$. BurstEngine demonstrates the capability to support training on extremely long sequences exceeding 1M tokens, paving the way for more efficient and scalable approaches to training more effective LLMs and LMMs.

\begin{acks}
This work is supported by the National Key Research and Development Program of China (2024YFB4505603), the National Natural Science Foundation of China (No.62192784), the high-quality development project of MIIT, and the Institute Guo Qiang at Tsinghua University. 
Cheng Yang is also supported by the Young Elite Scientists Sponsorship Program (No.2023QNRC001) by CAST.
\end{acks}

\newpage

\bibliographystyle{ACM-Reference-Format}
\bibliography{main}

\end{document}